\batchmode
\makeatletter
\def\input@path{{/Users/axel/Desktop/fBM/}}
\makeatother
\documentclass[english]{article}
\usepackage{lmodern}
\usepackage[T1]{fontenc}
\usepackage[latin9]{inputenc}
\usepackage[a4paper]{geometry}
\usepackage{amsmath}
\usepackage{amsthm}
\usepackage{amssymb}
\usepackage{graphicx}

\makeatletter
\newcommand{\lyxaddress}[1]{
	\par {\raggedright #1
	\vspace{1.4em}
	\noindent\par}
}
\newcommand\thmsname{\protect\theoremname}
\newcommand\nm@thmtype{theorem}
\theoremstyle{plain}

\newenvironment{namedthm}[1][Undefined Theorem Name]{
  \ifx{#1}{Undefined Theorem Name}\renewcommand\nm@thmtype{theorem*}
  \else\renewcommand\thmsname{#1}\renewcommand\nm@thmtype{namedtheorem}
  \fi
  \begin{\nm@thmtype}}
  {\end{\nm@thmtype}}

\usepackage{hyperref}
\usepackage[numbers,sort&compress]{natbib}
\date{}

\@ifundefined{showcaptionsetup}{}{%
 \PassOptionsToPackage{caption=false}{subfig}}
\usepackage{subfig}
\makeatother

\usepackage{babel}

\begin{document}
\title{\textbf{The fractional and mixed-fractional CEV model }}
\author{Axel A.~Araneda\thanks{Email: araneda@fias.uni-frankfurt.de, Tel.:+49 69 798 47501} }
\maketitle

\lyxaddress{\begin{center}
\vspace{-3em}Frankfurt Institute for Advanced Studies\\60438 Frankfurt
am Main, Germany.
\par\end{center}}

\begin{center}
\vspace{-1.5em} This version: June 1, 2019 \vspace{1.5em}
\par\end{center}
\begin{abstract}
The continuous observation of the financial markets has identified
some `stylized facts' which challenge the conventional assumptions,
promoting the born of new approaches. On the one hand, the long-range
dependence has been faced replacing the traditional Gauss-Wiener process
(Brownian motion), characterized by stationary independent increments,
by a fractional version. On the other hand, the CEV model addresses
the Leverage effect and smile-skew phenomena, efficiently. In this
paper, these two insights are merging and both the fractional and
mixed-fractional extensions for the CEV model, are developed. Using
the fractional versions of both the Itô's calculus and the Fokker-Planck
equation, the transition probability density function of the asset
price is obtained as the solution of a non-stationary Feller process
with time-varying coefficients, getting an analytical valuation formula
for a European Call option. Besides, the Greeks are computed and compared
with the standard case.


\vspace{2em}
\end{abstract}

\section{Introduction}

One of the most important insights in financial mathematics has been
the Black-Scholes model \cite{black1973pricing}, which uses a Geometric
Brownian motion (GBM) for describes the returns of the asset prices
as a regular diffusion process and arriving at an analytical formula
for a Vanilla European option. 

However, some ``Stylized facts'' in the financial markets don't agree
with the assumptions using in the Black-Scholes model. One of these
findings is the long-range dependence\footnote{a.k.a persistence, `Memory effect' or `Joseph effect'}
\cite{lo1991long,willinger1999stock,sadique2001long,cajueiro2004hurst}, motivating
the creation of a fractal version of the Black-Scholes model \cite{cutland1995stock,dai1996ito},
based on fractional Brownian motion \cite{mandelbrot1968fractional,mandelbrot2002gaussian,biagini2008stochastic}.
Hu \& \O skendal \cite{hu2003fractional} and Necula \cite{necula2002option}
arrive at an analytical formula for the European Call option for the
fractional Black-Scholes case, using Wick-Itô calculus \cite{duncan2000stochastic,nualart2008wick}.
Nonetheless, in the fractional framework, the arbitrage possibilities
aren't enterely omitted \cite{cheridito2003arbitrage,bjork2005note,rostek2013note}.
Addressing it fact, Cheridito \cite{cheridito2001mixed} introduces
the mixed-fractional Brownian motion (see further mathematical details
at refs. \cite{zili2006mixed,thale2009further}). This kind of model
ensures the absence of arbitrage opportunities \cite{androshchuk2006mixed,bender2007arbitrage}
and also a pricing formula for European type contracts could be obtained
\cite{sun2013pricing,murwaningtyas2018european}.

On the other hand, and come back to shortcomings of the original Black-Scholes
model, the homoscedasticity assumption is not consistent with other
empirical facts as the volatility smile-skew \cite{Derman1994,dupire1994pricing,rubinstein1985nonparametric,jackwerth1996recovering}
and leverage effect \cite{Black1976,christie1982stochastic,bekaert2000asymmetric,bollerslev2006leverage}.
The former is the change in the implied volatility pattern as a  function
of the strike price of an option. The latter is understood as the
inverse relationship between the volatility and the price. In this
context, a very popular formulation is the Constant Elasticity of
Variance (CEV) model developed by Cox \cite{Cox1975,cox1996constant},
which faces the heteroscedasticity and the leverage effect modeling
the volatility as a function of the asset price level. The model also
deals with the skew-smile phenomena \cite{beckers1980constant,chen2015equity}.
Despite that the CEV model considers only one more parameter (elasticity)
than the Black-Scholes model\footnote{See \cite{xiao2017estimating} to details on the parameter estimation
issue under CEV.}, the latter is outperformed by the CEV in both prices and option
pricing performance \cite{lee2004constant,macbeth1980tests,tucker1988tests,singh2011forecasting}.
Another plus point to the use of the CEV model is existence of a closed
form formula for a European vanilla option. The original Cox's work
derives the Call price in terms of summations of the incomplete gamma
function, but later Schroder \cite{schroder1989computing} developed
a closed-form solution depending on the non-central $\chi^2$ distribution\footnote{The computation of the non-central chi-squared distribution in Schroeder's
formula is quite unstable and becomes expensive for elasticities near
to zero. In order to reduce the computational times and also to address
American-type options, several analytical approximations and numerical
methods have been developed for the CEV model, see \cite{zhang2019multiquadric,larguinho2013computation}
for a survey.}. See Refs. \cite{wang2018optimal,yuan2018cev,yuan2019family} for
recent and successfully applications of the CEV model in different
contexts.

Given the previous statement, the aim of this paper is to merge the
local volatility approach and the fractional calculus, extending the
CEV model under classical Brownian motion to the fractional and mixed-fractional
cases. The fractional CEV case has been addressed previously in the
literature \cite{chan2006fractional}, proposing a European Call formula
in terms of the standard complementary gamma distribution function,
similar to the Cox's result, but without  explicit evaluation of the
added terms. This time, for the fractional CEV, the European Call
price is derived by a compact and explicit way, in terms of the non-central-chi-squared
distribution and the M-Whittaker function, following the Schroder
scheme and using a time-varying coefficients' version of the Feller's
diffusion problem. Besides, similarly, a pricing formula for the mixed-fractional
CEV model is studied. Also, the convergence of the fractional CEV
pricing to the fractional Black-Scholes case is shown. Moreover, the
Greeks of the models are computed and compared with the standard CEV
case.

The paper outline is the following. First, the CEV model is revisited.
Later, the fractional extension  is analyzed, deriving  the pricing
formula for a European Call option. After that, a mixed fractional
structure is proposed, arriving at the related European Call pricing
formula. At next, the computation and analysis of the Greeks, are
performed. Finally, the main conclusions are displayed.\\

\section{The CEV model\label{sec:The-CEV-model}}

Under the risk-neutral measure, at the constant elasticity of variance
model, the asset price $S$ follows the next stochastic differential
equation:\\

\begin{equation}
\mathrm{d}S=rS\mathrm{d}t+\sigma S^{\frac{\alpha}{2}}\mathrm{d}B_{t}\label{eq:CEV}
\end{equation}
\\

\noindent where $r,\,\sigma$ and $\alpha$ are the constant parameters
of the model, with $\sigma>0$ and $\alpha\in[0,2[$ . $B_{t}$ is
a standard Brownian motion, such that d$B_{t}\sim N\left(0,\mathrm{d}t\right)$.
In the limit case, when $\alpha\rightarrow2$, the CEV turns into
the Black-Scholes model. 

Aplying the following change of variable:

\begin{equation}
x(S,t)=S^{2-\alpha}\label{eq:cambio de variable}
\end{equation}
\\

\noindent and by the Itô's Lemma, Eq. (\ref{eq:CEV}) becomes:

\[
\mathrm{d}x=\left(2-\alpha\right)\left[rx+\frac{1}{2}\left(1-\alpha\right)\sigma^{2}\right]\mathrm{d}t+\left(2-\alpha\right)\sigma\sqrt{x}\mathrm{d}B_{t}
\]
\\

Let $P(x_{T},T|x_{0},0)$, the transition probability function which
rules the evolution of $x$ from $x(0)=x_{0}$ to $x(T)=x_{T}$, and
$T>0$. Then, $P$ evolves according to the Fokker-Planck equation:

\begin{equation}
\frac{\partial P}{\partial t}=\frac{1}{2}\frac{\partial^{2}}{\partial x^{2}}\left[\left(2-\alpha\right)^{2}\sigma^{2}xP\right]-\frac{\partial}{\partial x}\left[\left(2-\alpha\right)\left(rx+\frac{1}{2}\left(1-\alpha\right)\sigma^{2}\right)P\right]\label{eq:FP}
\end{equation}
\\

Eq. (\ref{eq:FP}) can be solved by the famous Feller's lemma \cite{feller1951two},
summarized at next:
\begin{namedthm}[Feller's lemma]
\hypertarget{Feller}{}

Let u=u(x,t), and a,b,c constants, with a>0 and t>0. The solution
of the parabolic equation

\[
\frac{\partial u}{\partial t}=\frac{\partial^{2}}{\partial x^{2}}\left[axu\right]-\frac{\partial}{\partial x}\left[\left(bx+c\right)u\right]
\]

\noindent conditional to

\[
u(x,0)=\delta(x-x_{0})
\]
\\

\noindent is given by

\[
u\left(\left.x,t\right|x_{0},0\right)=\frac{b}{a\left(\text{e}^{bt}-1\right)}\left(\frac{x\text{e}^{-bt}}{x_{0}}\right)^{{\textstyle {\displaystyle \frac{c-a}{2a}}}}\exp\left[-\frac{b\left(x+x_{0}\text{e}^{bt}\right)}{a\left(\text{e}^{bt}-1\right)}\right]I_{1-c/a}\left[\frac{2b}{a\left(1-\text{e}^{-bt}\right)}\sqrt{\text{e}^{-bt}x_{0}x}\right]
\]
\\

\noindent where $I_{k}(x)$ is the modified Bessel function of the
first kind of order k.
\end{namedthm}
\begin{proof}
See Refs. \cite{feller1951two,hsu2008constant}.
\end{proof}
If we set\label{a,b,c} $a=\left(2-\alpha\right)^{2}\sigma^{2}/2$,
$b=r\left(2-\alpha\right)$ and $c=\left(2-\alpha\right)\left(1-\alpha\right)\sigma^{2}/2$;
by the \hyperlink{Feller}{Feller's Lemma}, the transition probability
distribution from $x(0)=x_{0}$ to $x(T)=x_{T}$, is given by:

\begin{multline}
P\left(\left.x_{T},T\right|x_{0},0\right)=\frac{2r}{\sigma^{2}\left(2-\alpha\right)\left[\text{e}^{r\left(2-\alpha\right)T}-1\right]}\left(\frac{x_{T}}{x_{0}}\text{e}^{-r\left(2-\alpha\right)T}\right)^{{\textstyle -\frac{1}{2(2-\alpha)}}}\exp\left[-\frac{2r\left(x+x_{0}\text{e}^{r\left(2-\alpha\right)T}\right)}{\sigma^{2}\left(2-\alpha\right)\left(\text{e}^{r\left(2-\alpha\right)T}-1\right)}\right]\\
\times\,I_{1/\left(2-\alpha\right)}\left[\frac{4r}{\sigma^{2}\left(2-\alpha\right)\left(1-\text{e}^{-r\left(2-\alpha\right)T}\right)}\sqrt{\text{e}^{-r\left(2-\alpha\right)T}x_{0}x}\right]\label{eq:P(X_T)}
\end{multline}
\\

Coming back to the original variables and reordering terms, the density
for $S_{T}$ given $S_{0}$ is equal to \cite{schroder1989computing}:

\begin{eqnarray}
P\left(\left.S_{T},T\right|S_{0},0\right) & = & P\left(\left.x_{T},T\right|x_{0},0\right)\frac{\partial x_{T}}{\partial S_{T}}\nonumber \\
 & = & \left(2-\alpha\right)k^{\frac{1}{2-\alpha}}\left(yw^{1-2\alpha}\right)^{\frac{1}{2\left(2-\alpha\right)}}\text{e}^{-y-w}I_{1/\left(2-\alpha\right)}\left(2\sqrt{yw}\right)\label{eq:P}
\end{eqnarray}
\\

\noindent where:

\begin{eqnarray}
k & = & \frac{2r}{\sigma^{2}\left(2-\alpha\right)\left[\text{e}^{r\left(2-\alpha\right)T}-1\right]},\label{eq:k}\\
y & = & kS_{0}^{2-\alpha}\text{e}^{r\left(2-\alpha\right)T},\label{eq:y}\\
w & = & kS_{T}^{2-\alpha}.\label{eq:subst}
\end{eqnarray}
\\

Later, the value of a Call option at time $t=0$, with maturity $T$
and exercise price $E$, is computed by the Feynman-Kac formula:

\begin{eqnarray}
C\left(S,0\right) & = & \text{e}^{-rT}\int_{-\infty}^{\infty}\max\left\{ S_{T}-E,0\right\} P\left(\left.S_{T},T\right|S_{0},0\right)\text{d}S_{T}\nonumber \\
 & = & \text{e}^{-rT}\int_{E}^{\infty}\left(S_{T}-E\right)P\left(\left.S_{T},T\right|S_{0},0\right)\text{d}S_{T}\nonumber \\
 & = & \text{e}^{-rT}\int_{z}^{\infty}\left[\left(\frac{w}{k}\right)^{\frac{1}{2-\alpha}}-E\right]\left(2-\alpha\right)k^{\frac{1}{2-\alpha}}\left(yw^{1-2\alpha}\right)^{\frac{1}{2\left(2-\alpha\right)}}\text{e}^{-y-w}\nonumber \\
 &  & \times\,I_{1/\left(2-\alpha\right)}\left(2\sqrt{yw}\right)\left[\frac{1}{2-\alpha}\left(kw^{1-\alpha}\right)^{-\frac{1}{2-\alpha}}\right]\text{d}w\nonumber \\
 & = & \text{e}^{-rT}\int_{z}^{\infty}\left(\frac{w}{k}\right)^{\frac{1}{2-\alpha}}\left(\frac{y}{w}\right)^{\frac{1}{2\left(2-\alpha\right)}}\text{e}^{-y-w}I_{1/\left(2-\alpha\right)}\left(2\sqrt{yw}\right)\text{d}w\nonumber \\
 &  & -\text{e}^{-rT}\int_{z}^{\infty}E\left(\frac{y}{w}\right)^{\frac{1}{2\left(2-\alpha\right)}}\text{e}^{-y-w}I_{1/\left(2-\alpha\right)}\left(2\sqrt{yw}\right)\text{d}w\nonumber \\
 & = & C_{1}-C_{2}\label{eq:C_start}
\end{eqnarray}
\\

\noindent where $z=kE^{2-\alpha}$. As pointed by Schroder \cite{schroder1989computing},
the arguments of both integrals are the pdfs of the non-central chi-squared
distributions with $\nu$ degrees of freedom and non-centrality parameter
$\lambda$, noted by $\chi_{\nu}^{2}\left(\lambda\right)$ and defined
as:

\begin{eqnarray}
P_{\chi_{\nu}^{2}\left(\lambda\right)}(l) & = & \left(\frac{x}{\lambda}\right)^{\frac{\nu-2}{4}}\text{e}^{-\left(x+\lambda\right)/2}I_{\nu}\left(\sqrt{x\lambda}\right)\nonumber \\
 & = & f\left(l;\nu,\lambda\right)\label{eq:f}
\end{eqnarray}
\\

Back to the pricing equation, the first integral is developed as:

\begin{eqnarray*}
C_{1} & = & \text{e}^{-rT}\int_{z}^{\infty}k^{-\frac{1}{2-\alpha}}\left(wy\right)^{\frac{1}{2\left(2-\alpha\right)}}\text{e}^{-y-w}I_{1/\left(2-\alpha\right)}\left(2\sqrt{yw}\right)\text{d}w\\
 & = & \text{e}^{-rT}\int_{z}^{\infty}\left(\frac{y}{k}\right)^{\frac{1}{2-\alpha}}\left(w/y\right)^{\frac{1}{2\left(2-\alpha\right)}}\text{e}^{-y-w}I_{1/\left(2-\alpha\right)}\left(2\sqrt{yw}\right)\text{d}w\\
 & = & \text{e}^{-rT}\int_{z}^{\infty}\left(S_{0}\text{e}^{rT}\right)\left(w/y\right)^{\frac{1}{2\left(2-\alpha\right)}}\text{e}^{-y-w}I_{1/\left(2-\alpha\right)}\left(2\sqrt{yw}\right)\text{d}w\\
 & = & S_{0}\int_{z}^{\infty}\left(2w/2y\right)^{\frac{1}{2\left(2-\alpha\right)}}\text{e}^{-\left(2y+2w\right)/2}I_{2+\frac{2}{2-\alpha}}\left(\sqrt{\left(2y\right)\left(2w\right)}\right)\text{d}w\\
 & = & S_{0}\int_{z}^{\infty}f\left(2w,2+\frac{2}{2-\alpha},2y\right)\text{d}w\\
 & = & S_{0}\int_{z}^{\infty}f\left(2w,2+\frac{2}{2-\alpha},2y\right)\text{d}w
\end{eqnarray*}
\\

While the second one:

\begin{eqnarray*}
C_{2} & = & E\text{e}^{-rT}\int_{z}^{\infty}\left[\frac{2y}{2w}\right]^{\frac{1}{2\left(2-\alpha\right)}}\text{e}^{-\left(2y-2w\right)/2}I_{2+\frac{2}{2-\alpha}}\left(\sqrt{\left(2y\right)\left(2y\right)}\right)\text{d}w\\
 & = & E\text{e}^{-rT}\int_{z}^{\infty}f\left(2y,2+\frac{2}{2-\alpha},2w\right)\text{d}w
\end{eqnarray*}
\\

Called $Q$ to the complementary distribution function of $\chi_{\nu}^{2}\left(\lambda\right)$:

\[
\int_{m}^{\infty}f\left(l;\nu,\lambda\right)\text{d}l=Q\left(m,\nu,\lambda\right)
\]
\\

\noindent and using the following identity\cite{schroder1989computing}:

\[
\int_{m}^{\infty}f\left(2l;2\nu,2\lambda\right)\text{d}\lambda=1-Q\left(2l;2\nu-2,2m\right)
\]
\\

\noindent the call formula can be wrote as:

\begin{equation}
C\left(S,0\right)=S_{0}Q\left(2z;2+\frac{2}{2-\alpha},2y\right)-E\text{e}^{-rT}\left[1-Q\left(2y;\frac{2}{2-\alpha},2z\right)\right]\label{eq:SOL_CEV}
\end{equation}
\\

As we remarked previously, the Black-Scholes model can be treated
as a limit case of the CEV model, when $a\rightarrow2$. For observe
the convergence of the solution given in (\ref{eq:SOL_CEV}) to the
Black-Scholes case, we will use the following result for the complementary
distribution function $Q$, based on the central limit theorem \cite{horgan2013convergence}:

\begin{equation}
Q\left(m,\nu,\lambda\right)\approx Q_{N}\left(\frac{m-\left(\nu+\lambda\right)}{\sqrt{2\left(\nu+2\lambda\right)}}\right),\text{ as \ensuremath{\nu\rightarrow\infty} }\label{eq:CL}
\end{equation}
\\

\noindent  where $Q_{N}(\cdot)$ is the standard normal complementary
density function. 

Thus, the first complementary function of the Eq. (\ref{eq:SOL_CEV}),
when $\alpha\rightarrow2$, is computed as:

\begin{eqnarray}
\lim_{a\rightarrow2^{-}}Q\left(2z;2+\frac{2}{2-\alpha},2y\right) & = & Q_{N}\left[\lim_{a\rightarrow2^{-}}\frac{2z-2-\frac{2}{2-\alpha}-2y}{\sqrt{2\left(2+\frac{2}{2-\alpha}+4y\right)}}\right]\nonumber \\
 & = & Q_{N}\left[\lim_{a\rightarrow2^{-}}\frac{2rE^{2-\alpha}-2rS_{0}^{2-\alpha}\text{e}^{rT(2-\alpha)}+\sigma^{2}\left(3-\alpha\right)\left(\text{e}^{rT(2-\alpha)}-1\right)}{\sqrt{\sigma^{2}\left(2-\alpha\right)\left(\text{e}^{rT(2-\alpha)}-1\right)}}\right.\nonumber \\
 &  & \times\left.\frac{1}{\sqrt{4rS_{0}^{2-\alpha}\text{e}^{rT(2-\alpha)}+\sigma^{2}\left(3-\alpha\right)\left(\text{e}^{rT(2-\alpha)}-1\right)}}\right]\nonumber \\
 & = & Q_{N}\left[-\frac{\ln\left(\frac{S_{0}}{E}\right)+\left(r+\frac{1}{2}\sigma^{2}\right)T}{\sigma\sqrt{t}}\right]\nonumber \\
 & = & Q_{N}\left(-d_{1}\right)\label{eq:Q_d1}
\end{eqnarray}
\\

\noindent and for the symmetry of the normal function, we have that:

\[
Q_{N}\left(-d1\right)=N\left(d_{1}\right)
\]
\\

\noindent being $N(\cdot)$ the standard normal cumulative density.

The calculus of second $Q$ function of the Eq. (\ref{eq:SOL_CEV}),
when the degrees of freedom tends to infinity, is:

\begin{eqnarray}
\lim_{a\rightarrow2^{-}}Q\left(2y;\frac{2}{2-\alpha},2z\right) & = & Q_{N}\left[\lim_{a\rightarrow2^{-}}\frac{2y-\frac{2}{2-\alpha}-2x}{\sqrt{2\left(\frac{2}{2-\alpha}+4x\right)}}\right]\nonumber \\
 & = & Q_{N}\left[\lim_{a\rightarrow2^{-}}\frac{2rS_{0}^{2-\alpha}\text{e}^{rT(2-\alpha)}-2rE^{2-\alpha}-2\sigma^{2}\left(\text{e}^{rT(2-\alpha)}-1\right)}{\sqrt{\sigma^{2}\left(2-\alpha\right)\left(\text{e}^{rT(2-\alpha)}-1\right)}}\right.\nonumber \\
 &  & \times\left.\frac{1}{\sqrt{4rE^{2-\alpha}+\sigma^{2}\left(\text{e}^{rT(2-\alpha)}-1\right)}}\right]\nonumber \\
 & = & Q_{N}\left[\frac{\ln\left(\frac{S_{0}}{E}\right)+\left(r-\frac{1}{2}\sigma^{2}\right)T}{\sigma\sqrt{t}}\right]\nonumber \\
 & = & Q_{N}\left(d_{2}\right)\label{eq:Qd2}
\end{eqnarray}
\\

\noindent and for the identity between de cumulative and complementary
function:

\[
1-Q_{N}\left(d2\right)=N\left(d_{2}\right)
\]
\\

Later, at the limit $\alpha\rightarrow2$, the European Call pricing
of the CEV model converges to:

\[
\lim_{a\rightarrow2^{-}}C\left(S,0\right)=S_{0}N\left(d_{1}\right)-E\text{e}^{-rT}N\left(d_{2}\right)
\]
\\

\noindent which is the classical Black-Scholes formula provided in
\cite{black1973pricing}.

\section{A fractional CEV model}

Now, to address the `Joseph effect' at the CEV environment, the standard
Brownian motion of the Eq. (\ref{eq:CEV}) is switched by a fractional
one\footnote{Following \cite{duncan2000stochastic}, a fBM is a Gaussian process
which fulfills (for $0<H<1$; $t,s\geq0$):

i) $\mathbb{E}\left(B_{t}^{H}\right)=0$

ii) $\mathbb{E}\left(B_{t}^{H}\cdot B_{s}^{H}\right)=\frac{1}{2}\left\{ \left|t\right|^{2H}+\left|s\right|^{2H}-\left|t-s\right|^{2H}\right\} $ 

Then, for $H>1/2$, the autocorrelation function of $B_{t}^{H}$ is
positive and decays hyperbolically in function of the lags, i.e.,
long range dependency: $\sum_{n=1}^{\infty}\mathbb{E}\left[B_{1}^{H}\cdot\left(B_{n+1}^{H}-B_{n}^{H}\right)\right]=\infty$.}\textsuperscript{,}\footnote{Analogously to their classical counterpart, by the fractional Girsanov
theorem (see \cite{hu2003fractional,necula2002option,norros1999elementary}),
Eq. (\ref{eq:fCEV}) is wrote under the risk-neutral $\mathbb{Q}-$measure
with drift $r$, where $B_{H}^{t}$ is a $\mathbb{Q}-$fractional
Brownian motion. }\textsuperscript{,}\footnote{The fractional Brownian motion is not a semi-martingale for $H\neq1/2$;
i.e., there is not an equivalent martingale measure. As pointed in
\cite{sottinen2003arbitrage}, and despite the non-martingale condition,
the $\mathbb{Q-}$expected discounted value is equal to the current
value.}:

\begin{equation}
\mathrm{d}S=rS\mathrm{d}t+\sigma S^{\frac{\alpha}{2}}\mathrm{d}B_{t}^{H}\label{eq:fCEV}
\end{equation}
\\

\noindent where $B_{t}^{H}$ is a fractional Brownian motion with
Hurst exponent $H>1/2$.

Considering the shift of coordinates defined in Eq. (\ref{eq:cambio de variable})
and the fractional Itô's formula \cite{bender2003ito,biagini2004introduction},
the Eq. \ref{eq:fCEV} changes to:

\begin{equation}
\mathrm{d}x=\left(2-\alpha\right)\left[rx+Ht^{2H-1}\left(1-\alpha\right)\sigma^{2}\right]\mathrm{d}t+\left(2-\alpha\right)\sigma\sqrt{x}\mathrm{d}B_{t}\label{eq:dxf}
\end{equation}
\\

Then, the fractional Fokker-Planck equation \cite{metzler1999deriving,metzler2000random,unal2007fokker}
related to the stochastic process defined in Eq. (\ref{eq:dxf}),
is given by:

\begin{eqnarray}
\frac{\partial P_{H}}{\partial t} & = & \frac{\partial^{2}}{\partial x^{2}}\left[Ht^{2H-1}\left(2-\alpha\right)^{2}\sigma^{2}xP\right]-\frac{\partial}{\partial x}\left[\left(2-\alpha\right)\left(rx+Ht^{2H-1}\left(1-\alpha\right)\sigma^{2}\right)P\right]\nonumber \\
 & = & \frac{1}{2}\frac{\partial^{2}}{\partial x^{2}}\left[2Ht^{2H-1}\left(2-\alpha\right)^{2}\sigma^{2}xP\right]-\frac{\partial}{\partial x}\left[\left(2-\alpha\right)\left(rx+2Ht^{2H-1}\frac{1}{2}\left(1-\alpha\right)\sigma^{2}\right)P\right]\label{eq:fFP}
\end{eqnarray}
\\
Unfortunately, the relation (\ref{eq:fFP}) can't be solved using
the \hyperlink{Feller}{Feller's lemma} because the coefficient are
time-dependent (i.e, non-constant). However, Masoliver \cite{masoliver2016nonstationary}
provides an interesting approach for the non-stationary Feller process
when the coefficients are time-dependent, and the main useful result
for us is provided in the following statement:
\begin{namedthm}[Feller's lemma with time-varying coefficients]
\hypertarget{Feller_time}{}

Let u=u(x,$\tau$), A=A($\tau$), C=C($\tau$) and $\theta$ constant
defined by

\[
\theta=\frac{C(\tau)}{A(\tau)}
\]

The solution of the parabolic equation

\[
\frac{\partial u}{\partial\tau}=\frac{\partial^{2}}{\partial x^{2}}\left[Axu\right]+\frac{\partial}{\partial x}\left[\left(x-C\right)u\right]
\]
\\

\noindent conditional to

\[
u(x,0)=\delta(x-x_{0})
\]
\\

\noindent is given by

\[
u\left(\left.x,\tau\right|x_{0},0\right)=\frac{1}{\phi(\tau)}\left(\frac{x\text{e}^{\tau}}{x_{0}}\right)^{{\textstyle {\displaystyle \frac{\theta-1}{2}}}}\exp\left[-\frac{\left(x+x_{0}\text{e}^{-\tau}\right)}{\phi(\tau)}\right]I_{1-\theta}\left[\frac{2}{\phi(\tau)}\sqrt{\text{e}^{\tau}x_{0}x}\right]
\]
\\

\noindent where $I_{k}(x)$ is the modified Bessel function of the
first kind of order k and
\[
\phi(\tau)=\int_{0}^{\tau}A(\tau-s)\text{e}^{-s}\text{d}s
\]
\end{namedthm}
\begin{proof}
See Ref. \cite{masoliver2016nonstationary}.\\
\end{proof}
So, if we use the previous definitions of $a,b,c$ (cf. page \pageref{a,b,c}),
and set:

\begin{eqnarray}
\tau & = & -bt\label{eq:tau}\\
A(\tau) & = & -\frac{a}{b}2H\left(-\frac{\tau}{b}\right)^{2H-1}\label{eq:A}\\
C(\tau) & = & -\frac{c}{b}2H\left(-\frac{\tau}{b}\right)^{2H-1}\label{eq:C}
\end{eqnarray}

\noindent   Eq. (\ref{eq:fFP}) transforms into:

\begin{equation}
\frac{\partial P_{H}}{\partial\tau}=\frac{\partial^{2}}{\partial x^{2}}\left[A(\tau)xP\right]+\frac{\partial}{\partial x}\left[\left(x+C\left(\tau\right)\right)P\right]\label{eq:fFP_TV}
\end{equation}

\noindent and $\theta=C\left(\tau\right)/A\left(\tau\right)=c/a$,
a constant.

Then, Eq. (\ref{eq:fFP_TV}), can be solved by the \hyperlink{Feller_time}{Feller's lemma with time-varying coefficients}.
Indeed:

\[
P_{H}\left(\left.x,\tau\right|x_{0},0\right)=\frac{1}{\phi(\tau)}\left(\frac{x\text{e}^{\tau}}{x_{0}}\right)^{{\textstyle {\displaystyle \frac{c-a}{2a}}}}\exp\left[-\frac{\left(x+x_{0}\text{e}^{-\tau}\right)}{\phi(\tau)}\right]I_{1-c/a}\left[\frac{2}{\phi(\tau)}\sqrt{\text{e}^{-\tau}x_{0}x}\right]
\]
\\
\noindent where:

\begin{eqnarray*}
\phi(\tau) & = & -\frac{a}{b}\int_{0}^{\tau}2H\left(\frac{s-\tau}{b}\right){}^{2H-1}\text{e}^{-s}\text{d}s\\
 & = & \frac{a}{2H+1}\left(-\frac{\tau}{b}\right)^{2H}\left[2H+1+\text{e}^{-\frac{1}{2}\tau}\left(-\tau\right)^{-H}M_{H,H+1/2}\left(-\tau\right)\right]
\end{eqnarray*}
 \\

\noindent and $M_{\kappa,\upsilon}\left(l\right)$ is the M-Whittaker
function\footnote{The Whittaker (or  confluent hypergeometric) function appears in the
solution of other related problems that involve the CEV process, see
for example \cite{albanese2001black,mendoza2011pricing,jo2016convergence},
among others.} \cite{whittaker1903expression,buchholz_confluent,NIST} and can be
expressed in terms of the $M$ confluent hypergeometric Kummer's function:

\[
M_{\kappa,\upsilon}\left(l\right)=l^{\upsilon+1/2}\text{e}^{-l/2}M\left(\upsilon-\kappa+\frac{1}{2},1+2\upsilon;l\right)
\]
\\

Now, solving for the original time-coordinate, at time $t=T$, we
have:

\[
P_{H}\left(\left.x,T\right|x_{0},0\right)=\frac{1}{\phi(T)}\left(\frac{x\text{e}^{-bT}}{x_{0}}\right)^{{\textstyle {\displaystyle \frac{c-a}{2a}}}}\exp\left[-\frac{\left(x+x_{0}\text{e}^{bT}\right)}{\phi(\tau)}\right]I_{1-c/a}\left[\frac{2}{\phi(T)}\sqrt{\text{e}^{bT}x_{0}x}\right]
\]
\\

\noindent with

\begin{eqnarray*}
\phi(T) & = & \frac{a}{2H+1}T^{2H}\left[2H+1+\text{e}^{\frac{1}{2}bT}\left(bT\right)^{-H}M_{H,H+1/2}\left(bT\right)\right]
\end{eqnarray*}
\\

Later, moving to the original frame of reference $(S,t)$, and replacing
the values for $a,$$b$ and $c$, the probability density function
of $S(T)=S_{T}$, $T>0,$ given $S(0)=S_{0}$ is:

\begin{eqnarray}
P_{H}\left(\left.S_{T},T\right|S_{0},0\right) & = & P_{H}\left(\left.x_{T},T\right|x_{0},0\right)\frac{\partial x_{T}}{\partial S_{T}}\nonumber \\
 & = & \left(2-\alpha\right)k_{H}^{\frac{1}{2-\alpha}}\left(yw^{1-2\alpha}\right)^{\frac{1}{2\left(2-\alpha\right)}}\text{e}^{-y-w}I_{1/\left(2-\alpha\right)}\left(2\sqrt{yw}\right)\label{eq:PH}
\end{eqnarray}
\\

\noindent being:

\begin{eqnarray}
k_{H} & = & \left[\phi\left(T\right)\right]^{-1},\label{eq:k-1}\\
y_{H} & = & k_{H}S_{0}^{2-\alpha}\text{e}^{r\left(2-\alpha\right)T}\\
w_{H} & = & k_{H}S_{T}^{2-\alpha}
\end{eqnarray}

The transition probabilities $P$ (Eq. (\ref{eq:P})) and $P_{H}$
(Eq. (\ref{eq:PH})) differ only by the terms $k$ and $k_{H}$. For
the particular case $H=1/2$, these terms are equal\footnote{
\begin{eqnarray*}
M_{1/2,1}\left(l\right)=\frac{2\text{e}^{-l/2}\left[\text{e}^{l}-\left(l+1\right)\right]}{\sqrt{l}} & \implies & \text{e}^{l/2}\left(l\right)^{-1/2}M_{1/2,1}\left(l\right)=\frac{2}{l}\left[\text{e}^{l}-\left(l+1\right)\right]\\
 & \implies & \frac{l}{2b}\left[2+\text{e}^{l/2}\left(l\right)^{-1/2}M_{1/2,1}\left(l\right)\right]=\frac{1}{b}\left(\text{e}^{l}-1\right)\\
 & \implies & \phi(t)\biggr|_{H=1/2}=\frac{a}{b}\left(\text{e}^{bt}-1\right)\\
 & \implies & k_{H}\biggr|_{H=1/2}=\frac{b}{a\left(\text{e}^{bT}-1\right)}=k
\end{eqnarray*}
}, yield to $P=P_{H}\Bigr|_{H=1/2}$ .

After that, a European option price may be computed taking expectations
of the discounted payoff (see appendix \ref{sec:Risk-neutral-pricing-in}).
Fixing ${\displaystyle z_{H}(t)=k_{H}(t)E^{2-\alpha}}$, and following
the development given from the Eq.(\ref{eq:C_start}) to the Eq. (\ref{eq:SOL_CEV}),
the pricing for a European call , in the fractal CEV, becomes:

\begin{equation}
C_{H}\left(S_{0},0\right)=S_{0}Q\left(2z_{H};2+\frac{2}{2-\alpha},2y_{H}\right)-E\text{e}^{-rT}\left[1-Q\left(2y_{H};\frac{2}{2-\alpha},2z_{H}\right)\right]\label{eq:SOL_CEV-1}
\end{equation}
\\
Using the derivatives, asymptotics and recurrence properties\footnote{Mainly, we use:
\begin{itemize}
\item ${\displaystyle \lim_{l\rightarrow0}M_{\kappa,\upsilon}\left(l\right)=0}$
\item ${\displaystyle \frac{\text{\ensuremath{\partial}}}{\partial l}}M_{\kappa,\upsilon}\left(l\right)=\left(\frac{1}{2}-\frac{\kappa}{l}\right)M_{\kappa,\upsilon}\left(l\right)+l^{-1}\left(\frac{1}{2}+\kappa+\upsilon\right)M_{\kappa+1,\upsilon}\left(l\right)$
\item $M_{\kappa,\kappa+1/2}\left(l\right)+M_{\kappa+1,\kappa+1/2}\left(l\right)=M_{\kappa,\kappa+1/2}\left(l\right)+l^{\kappa+1}\text{e}^{-l/2}$
\end{itemize}
} of both $M_{\kappa,\upsilon}\left(l\right)$ and $M$ \cite{NIST},
from (\ref{eq:SOL_CEV-1}) an interesting result is obtained computing
the limit case $\alpha\rightarrow2$. By Eq. \ref{eq:CL}, we get:

\begin{eqnarray}
\lim_{\alpha\rightarrow2^{-}}Q\left(2z_{H};2+\frac{2}{2-\alpha},2y_{H}\right) & = & Q_{N}\left[\lim_{a\rightarrow2^{-}}\frac{2z_{H}-2-\frac{2}{2-\alpha}-2y_{H}}{\sqrt{2\left(2+\frac{2}{2-\alpha}+4y_{H}\right)}}\right]\nonumber \\
 & = & Q_{N}\left[\lim_{a\rightarrow2^{-}}\frac{E^{2-alpha}-\phi(T)\left(1+\frac{1}{2-\alpha}\right)-S_{0}^{2-\alpha}\text{e}^{(2-\alpha)T}}{\sqrt{\phi(T)}\sqrt{\phi(T)\left(1+\frac{1}{2-\alpha}\right)+2S_{0}^{2-\alpha}\text{e}^{(2-\alpha)T}}}\right]\nonumber \\
 & = & Q_{N}\left[-\frac{\ln\left(\frac{S_{0}}{E}\right)+rT+\frac{1}{2}\sigma^{2}T^{2H}}{\sigma\sqrt{T^{2H}}}\right]\nonumber \\
 & = & Q_{N}\left(-d_{1}^{H}\right)\nonumber \\
 & = & N\left(d_{1}^{H}\right)\label{eq:ND1H}
\end{eqnarray}
\\
\begin{eqnarray}
\lim_{\alpha\rightarrow2^{-}}Q\left(2y_{H};\frac{2}{2-\alpha},2z_{H}\right) & = & Q_{N}\left[\lim_{a\rightarrow2^{-}}\frac{2y_{H}-\frac{2}{2-\alpha}-2z_{H}}{\sqrt{2\left(\frac{2}{2-\alpha}+4z_{H}\right)}}\right]\nonumber \\
 & = & Q_{N}\left[\lim_{a\rightarrow2^{-}}\frac{S_{0}^{2-\alpha}\text{e}^{(2-\alpha)T}-E^{2-alpha}-\phi(T)\left(\frac{1}{2-\alpha}\right)}{\sqrt{\phi(T)}\sqrt{\phi(T)\left(\frac{1}{2-\alpha}\right)+2E_{0}^{2-\alpha}}}\right]\nonumber \\
 & = & Q_{N}\left[\frac{\ln\left(\frac{S_{0}}{E}\right)+rT-\frac{1}{2}\sigma^{2}T^{2H}}{\sigma\sqrt{T^{2H}}}\right]\nonumber \\
 & = & Q_{N}\left(d_{2}^{H}\right)\nonumber \\
 & = & 1-N\left(d_{2}^{H}\right)\label{eq:ND2H}
\end{eqnarray}
\\
\noindent and replacing in (\ref{eq:SOL_CEV-1}), we arrive at the
fractional Black-Scholes formula \cite{hu2003fractional,necula2002option}:

\[
C_{H}\left(S,0\right)=S_{0}N\left(d_{1}^{H}\right)-E\text{e}^{-rT}N\left(d_{2}^{H}\right)
\]
\\

Then, the convergence of the CEV to the Black-Scholes model, in the
limit case $\alpha\rightarrow2,$ remains in the fractional scheme.

\begin{figure}
\subfloat[\label{fig:T=00003D0.25}T=0.25]%
{\includegraphics[width=0.5\textwidth]{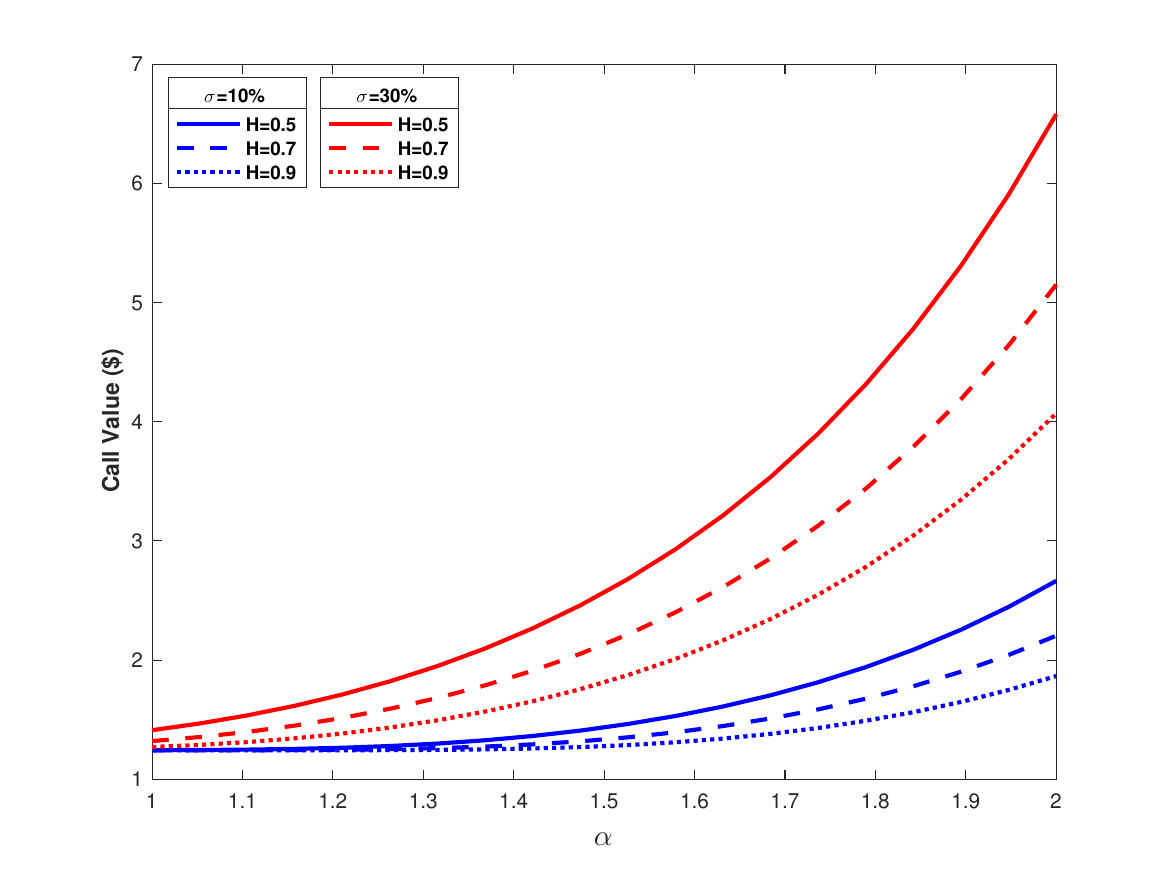}

}%
\subfloat[\label{fig:T=00003D0.5}T=0.5]%
{\includegraphics[width=0.5\textwidth]{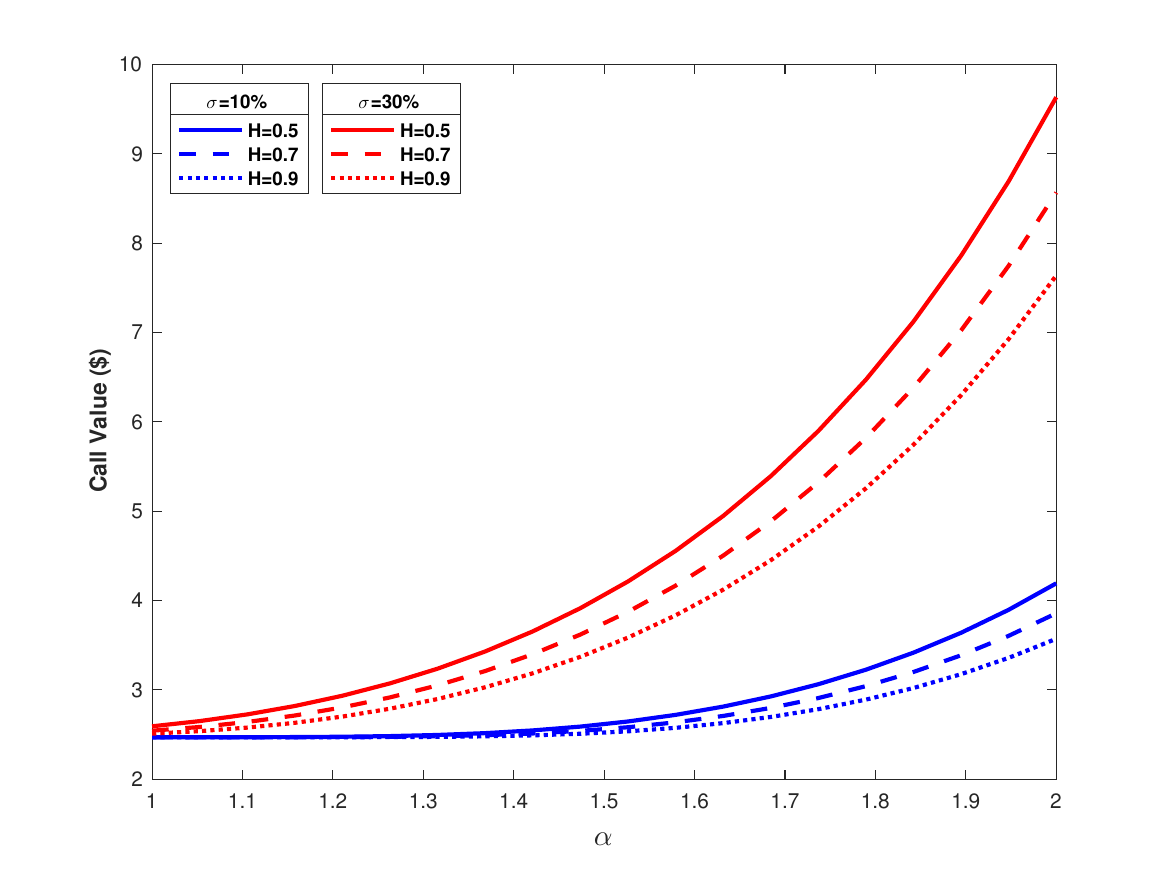}

}\\
\subfloat[\label{fig:T=00003D1.5}T=1.5]%
{\includegraphics[width=0.5\textwidth]{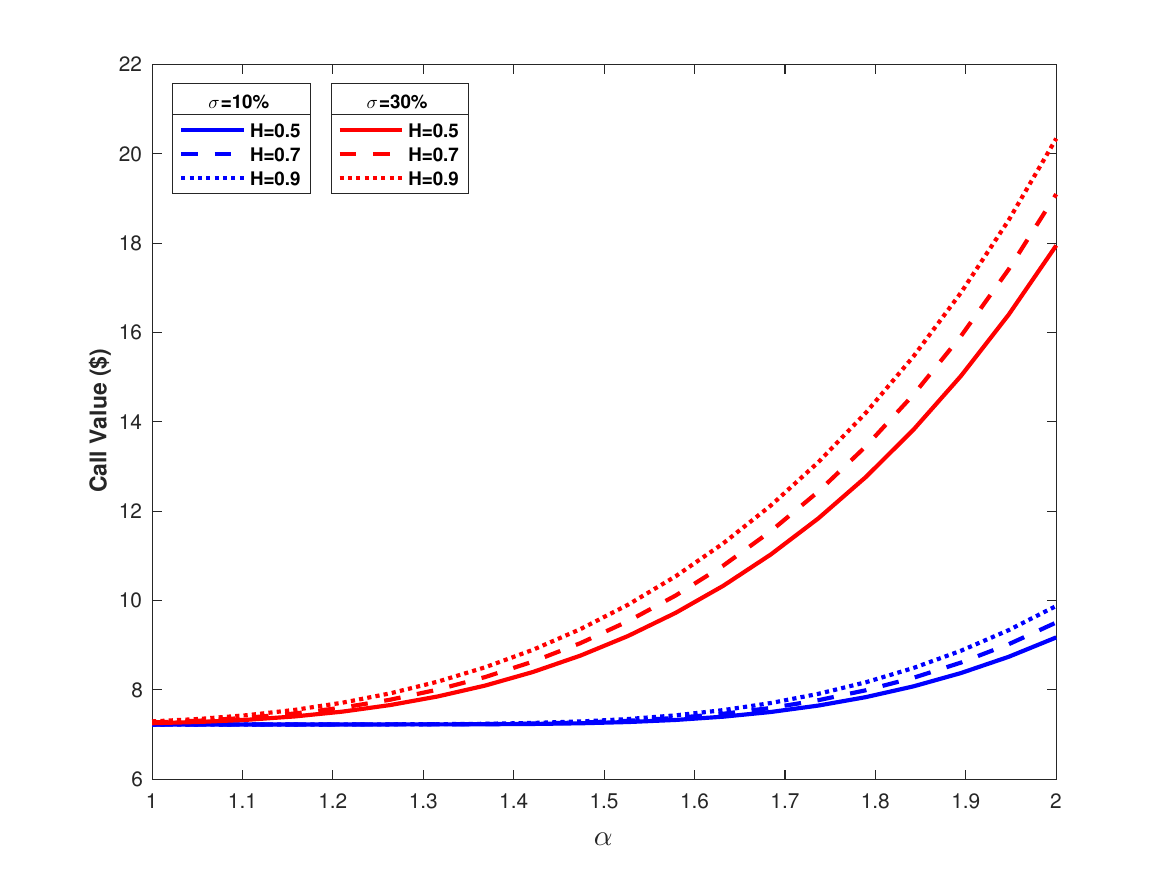}

}%
\subfloat[\label{fig:T=00003D2}T=2]%
{\includegraphics[width=0.5\textwidth]{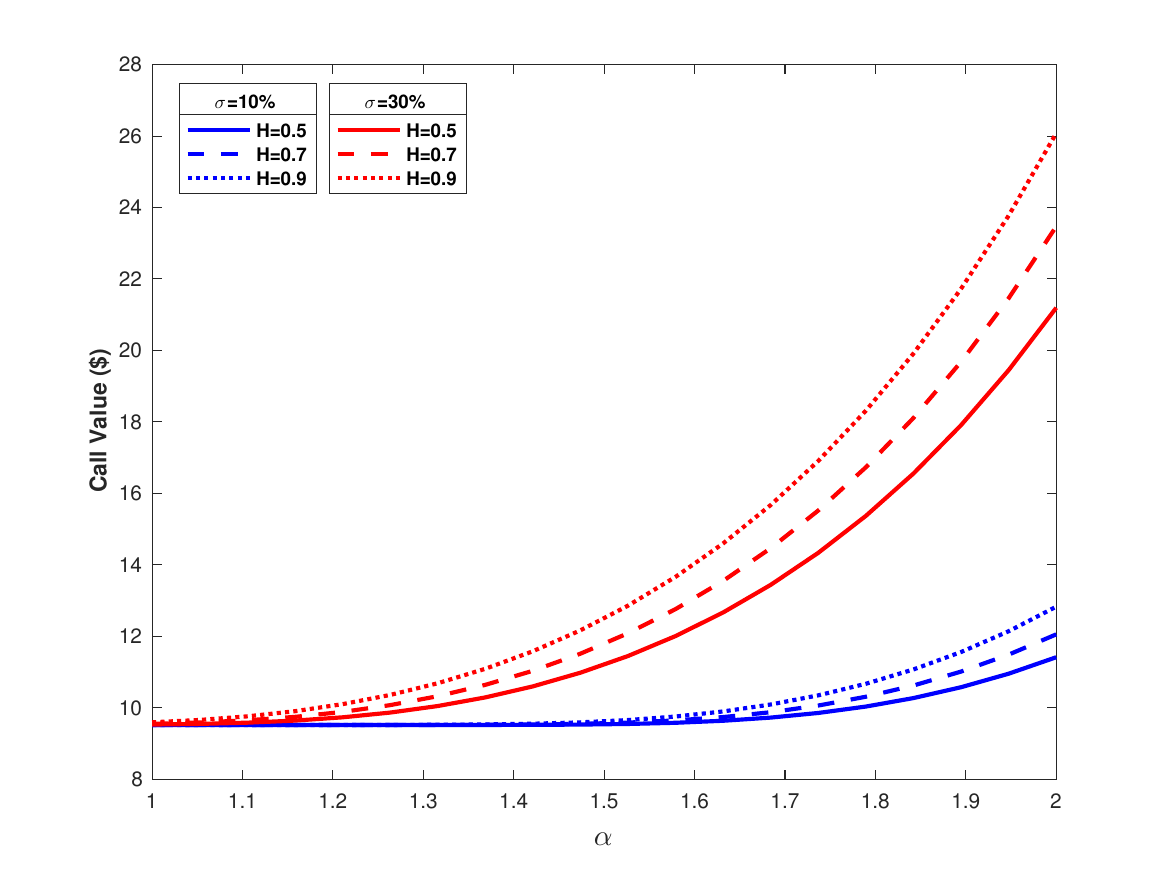}

}

\caption{\label{fig:Fractional-CEV-formula}Fractional CEV formula for a European
Call with $\sigma$=\{10\%; 30\%\} and $H=$\{0.5; 0.7; 0.9\} and
different maturities in function of the elasticity parameter, fixing
$S_{0}=E=100$ and $r=5$\%. }
\end{figure}

Fig. \ref{fig:Fractional-CEV-formula} plots the values of the fractional
CEV formula for $\sigma$ equal to 15\% (blue) and 30\% (red), and
three values of $H$=\{0.5, 0.7, 0.9\} considering different maturities.
The solid lines draw the case $H=1/2$, which corresponds to the classical
CEV pricing. The semi-solid and dotted lines show the pricing using
$H=0.7$ and $H=0.9$, respectively. The fractional CEV retains the
property of being a monotonically increasing function of the elasticity
parameter. Also, is a rising function of $\sigma$ and $T$. For expiration
times below the year (Figs. \ref{fig:T=00003D0.25}-\ref{fig:T=00003D0.5}),
the option price falls if $H$ moves to 1. In the opposite way, for
$T>1$ (Figs. \ref{fig:T=00003D1.5}-\ref{fig:T=00003D2}), the prices
grow if $H$ rise in the interval {[}1/2,1{[}. When $\alpha=2$, the
fractional CEV pricing transforms into the fractional Black-Scholes
price.

\section{A mixed-fractional CEV model}

A mixed-fractional Brownian motion, is defined as a linear combination
between  an standard Brownian motion and other independent and fractional
Brownian motion\footnote{Cheridito \cite{cheridito2001mixed} proves that for $H\in]3/4,1[$
the filtration generated by $M_{t}^{\beta,\gamma,H}$ is equivalent
to a classical Brownian motion; i.e, a semi-martingale.}:

\[
M_{t}^{\beta,\gamma,H}=\beta B_{t}+\gamma B_{t}^{H};\qquad\beta\geq0,\,\gamma\geq0,
\]
\\

Then, for to extent the CEV model to the mixed-fractional case, the
Brownian motion which drives the CEV model is replaced by $M_{t}^{H,\beta}$:

\begin{eqnarray}
\mathrm{d}S & = & rS\mathrm{d}t+\sigma S^{\frac{\alpha}{2}}\mathrm{d}M_{t}^{\beta,\gamma,H}\nonumber \\
 & = & rS\mathrm{d}t+\sigma S^{\frac{\alpha}{2}}\left(\beta\mathrm{d}B_{t}+\gamma\mathrm{d}B_{t}^{H}\right)\label{eq:mfCEV}
\end{eqnarray}
\\
Is clear that if $\left(\beta,\gamma\right)=\left(0,1\right)$ we
recover the fractional case studied in the previous section. Also,
if $\beta=1$ and $\gamma=0$,  Eq. (\ref{eq:mfCEV}) describes the
classical CEV model (Sec. \ref{sec:The-CEV-model}). 

Analogous to the previous cases, the transformation (\ref{eq:cambio de variable})
and the fractional Itô's lemma goes  Eq. (\ref{eq:mfCEV}) to:

\[
\mathrm{d}x=\left(2-\alpha\right)\left[rx+\left(\frac{1}{2}\beta+\gamma Ht^{2H-1}\right)\left(1-\alpha\right)\sigma^{2}\right]\mathrm{d}t+\left(2-\alpha\right)\sigma\sqrt{x}\left(\beta\mathrm{d}B_{t}+\gamma\mathrm{d}B_{t}^{H}\right)
\]
\\

Since that $B_{t}=B_{t}^{1/2}$, the fractional Fokker-Planck equation
for the above process is:

\begin{equation}
\frac{\partial P_{M}}{\partial t}=\frac{1}{2}\frac{\partial^{2}}{\partial x^{2}}\left[\left(\beta+2\gamma Ht^{2H-1}\right)\left(2-\alpha\right)^{2}\sigma^{2}xP\right]-\frac{\partial}{\partial x}\left\{ \left(2-\alpha\right)\left[rx+\left(\beta+2\gamma Ht^{2H-1}\right)\frac{\left(1-\alpha\right)}{2}\sigma^{2}\right]P\right\} \label{eq:MF-FPE}
\end{equation}
\\
Setting:

\begin{eqnarray*}
A'(\tau) & = & \gamma A(\tau)-\frac{a}{b}\beta\\
C'(\tau) & = & \gamma C(\tau)-\frac{a}{b}\beta\\
\theta' & = & \frac{C'(\tau)}{A'(\tau)}
\end{eqnarray*}

\noindent the relation (\ref{eq:MF-FPE}) becomes:

\begin{equation}
\frac{\partial P_{M}}{\partial\tau}=\frac{\partial^{2}}{\partial x^{2}}\left[A'(\tau)xP\right]+\frac{\partial}{\partial x}\left\{ \left(2-\alpha\right)\left[x-C'(\tau)\right]P\right\} \label{eq:MFFP_2}
\end{equation}

\noindent where $\tau$ $A$, $C$ are given by Eqs. (\ref{eq:tau})-(\ref{eq:C}). 

Given that $\theta'=\theta=c/a$ (constant), a solution for (\ref{eq:MFFP_2})
is obtained  through the \hyperlink{Feller_time}{Feller's lemma with time-varying coefficients}:

\[
P_{M}\left(\left.x,\tau\right|x_{0},0\right)=\frac{1}{\phi\text{'}(\tau)}\left(\frac{x\text{e}^{\tau}}{x_{0}}\right)^{{\textstyle {\displaystyle \frac{c-a}{2a}}}}\exp\left[-\frac{\left(x+x_{0}\text{e}^{-\tau}\right)}{\phi'(\tau)}\right]I_{1-c/a}\left[\frac{2}{\phi'(\tau)}\sqrt{\text{e}^{-\tau}x_{0}x}\right]
\]

and

\begin{eqnarray*}
\phi(\tau) & = & -\frac{a}{b}\int_{0}^{\tau}\gamma\left[2H\left(\frac{s-\tau}{b}\right){}^{2H-1}-\beta\right]\text{e}^{-s}\text{d}s\\
 & = & \gamma\frac{a}{2H+1}\left(-\frac{\tau}{b}\right)^{2H}\left[2H+1+\text{e}^{-\frac{1}{2}\tau}\left(-\tau\right)^{-H}M_{H,H+1/2}\left(-\tau\right)\right]+\beta\frac{a}{b}\left(\text{e}^{-\tau}-1\right)
\end{eqnarray*}
\\

Later, the transition probability density function for $S_{T}$ conditional
to $S_{0}$, under a mixed-fractional regime is given by:

\begin{eqnarray*}
P_{M}\left(\left.S_{T},T\right|S_{0},0\right) & = & \left(2-\alpha\right)k_{M}^{\frac{1}{2-\alpha}}\left(yw^{1-2\alpha}\right)^{\frac{1}{2\left(2-\alpha\right)}}\text{e}^{-y-w}I_{1/\left(2-\alpha\right)}\left(2\sqrt{yw}\right)
\end{eqnarray*}
\\

\noindent being :

\begin{eqnarray}
k_{M} & = & \left[\phi'\left(T\right)\right]^{-1},\label{eq:k-1-1}\\
y_{M} & = & k_{M}S_{0}^{2-\alpha}\text{e}^{r\left(2-\alpha\right)T}\\
w_{M} & = & k_{M}S_{T}^{2-\alpha}
\end{eqnarray}
\\

\noindent with,

\begin{eqnarray*}
\phi'(T) & = & \frac{a}{2H+1}T^{2H}\left[2H+1+\text{e}^{\frac{1}{2}bT}\left(bT\right)^{-H}M_{H,H+1/2}\left(bT\right)\right]+\beta\frac{a}{b}\left(\text{e}^{-\tau}-1\right)\\
 & = & \gamma\phi(T)+\beta\frac{a}{b}\left(\text{e}^{-\tau}-1\right)
\end{eqnarray*}
\\

Using the argument provided in the previous sections, the European
Call price, at time $t=0$ under the mixed-fractional CEV framework
is computed by:

\begin{equation}
C_{M}\left(S_{0},0\right)=S_{0}Q\left(2z_{M};2+\frac{2}{2-\alpha},2y_{M}\right)-E\text{e}^{-rT}\left[1-Q\left(2y_{M};\frac{2}{2-\alpha},2z_{M}\right)\right]\label{eq:SOL_CEV-1-1}
\end{equation}
\\
\noindent where ${\displaystyle z_{H}(t)=k_{H}(t)E^{2-\alpha}}$.

Eq. (\ref{eq:SOL_CEV-1-1}) turns into the pure fractional CEV case
(Eq. (\ref{eq:SOL_CEV-1})) if $\left(\beta,\gamma\right)=\left(1,0\right)$,
and becomes the classical CEV model if $\left(\beta,\gamma\right)=\left(1,0\right)$
or $\left(\beta,\gamma,H\right)=\left(0,1,1/2\right)$.

Using the results computed in the Eqs. (\ref{eq:Q_d1}), (\ref{eq:Qd2}),
(\ref{eq:ND1H}) and (\ref{eq:ND2H}); at the limit case $\alpha\rightarrow2,$
Eq.(\ref{eq:SOL_CEV-1-1}) tends to:

\[
\lim_{\alpha\rightarrow2^{-}}C_{M}\left(S_{0},0\right)=S_{0}N\left(d_{1}^{M}\right)-E\text{e}^{-rT}N\left(d_{2}^{M}\right)
\]
\\
\noindent with

\[
d_{1}^{M}=\frac{\ln\left(\frac{S_{0}}{E}\right)+rT+\frac{1}{2}\beta\sigma^{2}T+\frac{1}{2}\gamma\sigma^{2}T^{2H}}{\sigma\sqrt{\beta T+\gamma T^{2H}}}
\]
\\

\[
d_{2}^{M}=\frac{\ln\left(\frac{S_{0}}{E}\right)+rT-\frac{1}{2}\beta\sigma^{2}T-\frac{1}{2}\gamma\sigma^{2}T^{2H}}{\sigma\sqrt{\beta T+\gamma T^{2H}}}
\]
\\
\noindent which is the mixed-fractional Black-Scholes pricing formula
(for instance, see \cite{murwaningtyas2018european} with $\beta=1$
and $\gamma=1$); keeping the convergence between the CEV and Black-Scholes.

\begin{figure}
\subfloat[\label{fig:T=00003D0.5-1}T=0.5]%
{\includegraphics[width=0.5\textwidth]{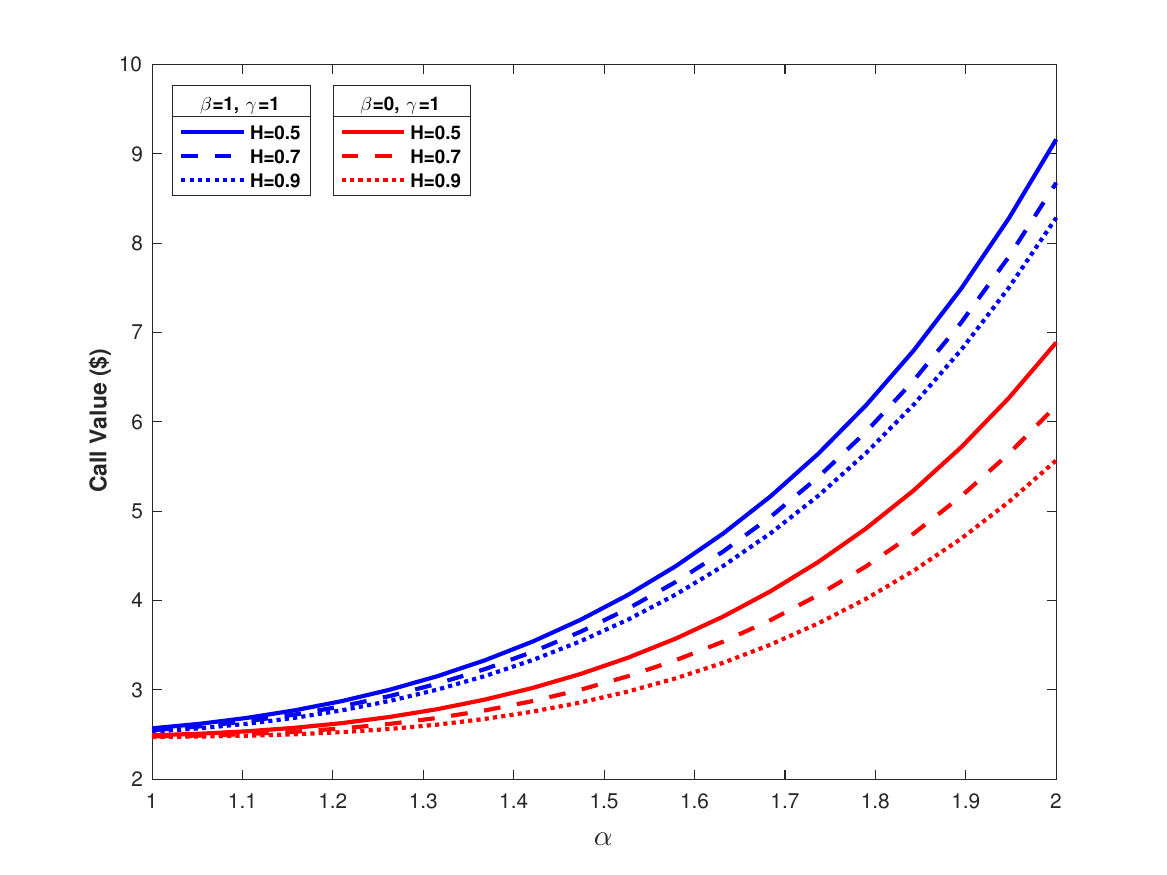}

}%
\subfloat[\label{fig:T=00003D1.5-1}T=1.5]%
{\includegraphics[width=0.5\textwidth]{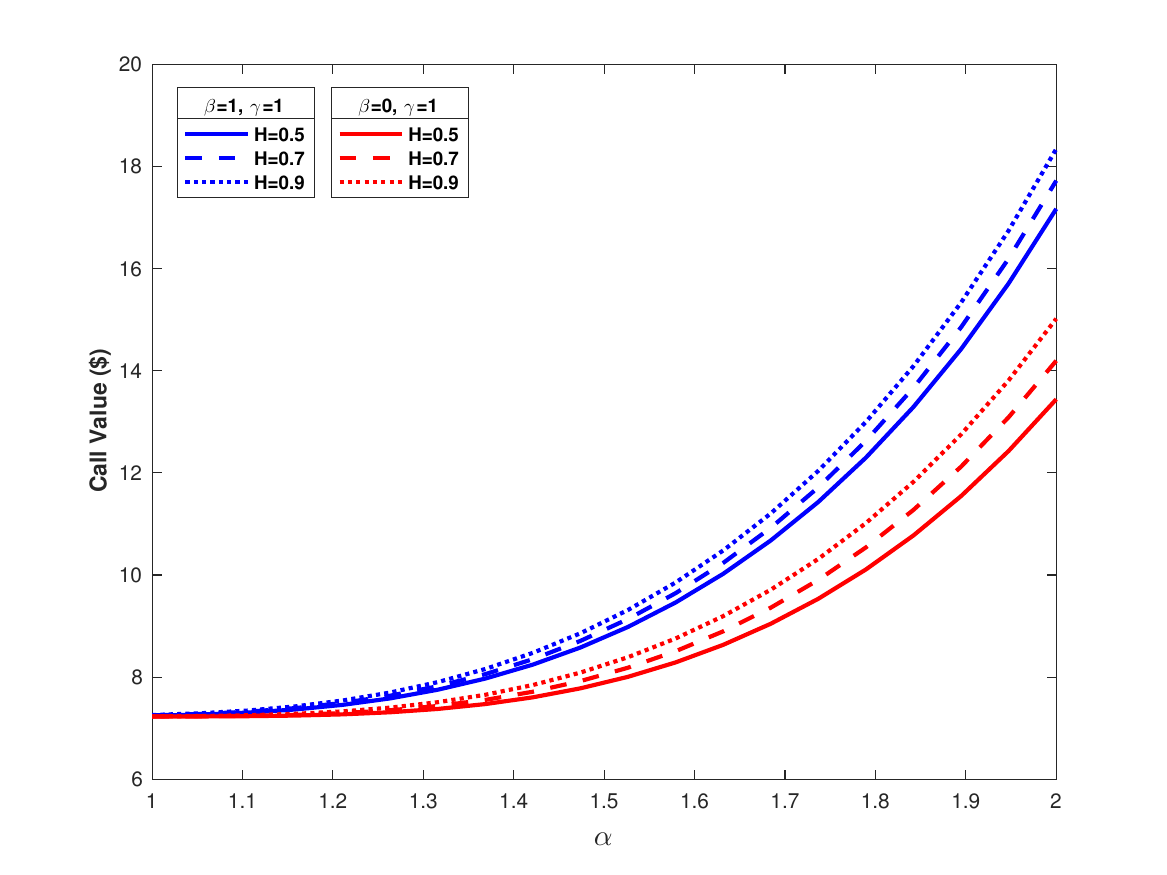}

}

\caption{\label{fig:Mixed-fractional-and-fractional}Mixed-fractional and fractional
CEV formula using $S_{0}=E=100$, $r=5$\% and $\sigma=20$\%}
\end{figure}

Fig. \ref{fig:Mixed-fractional-and-fractional} display the value
for a European Call under the mixed-fractional CEV model, setting
the pair of coefficients $\beta$ and $\gamma$ as $\left(\beta,\gamma\right)=\left(1,1\right)$
and $\left(\beta,\gamma\right)=\left(0,1\right)$ (blue and red respectively),
with the aim of to compare the mixed fractional and pure fractional
cases. We consider two maturities T=0.5 (\ref{fig:T=00003D0.5-1})
and T=1.5 (\ref{fig:T=00003D1.5-1}) and $H\in\left\{ 0.5;0.7;0.9\right\} $.
In both subplots the classical CEV pricing is represented by the red-solid-line.
We can observe that the mixed-fractional price is higher than the
both classical and fractional price. As in the pure fractional case,
for T<1 the price decreases if H tends to one, and increases for T>1
and H moves to one. 

\begin{figure}
\subfloat[\label{fig:T=00003D0.5-1-1}T=0.5]%
{\includegraphics[width=0.5\textwidth]{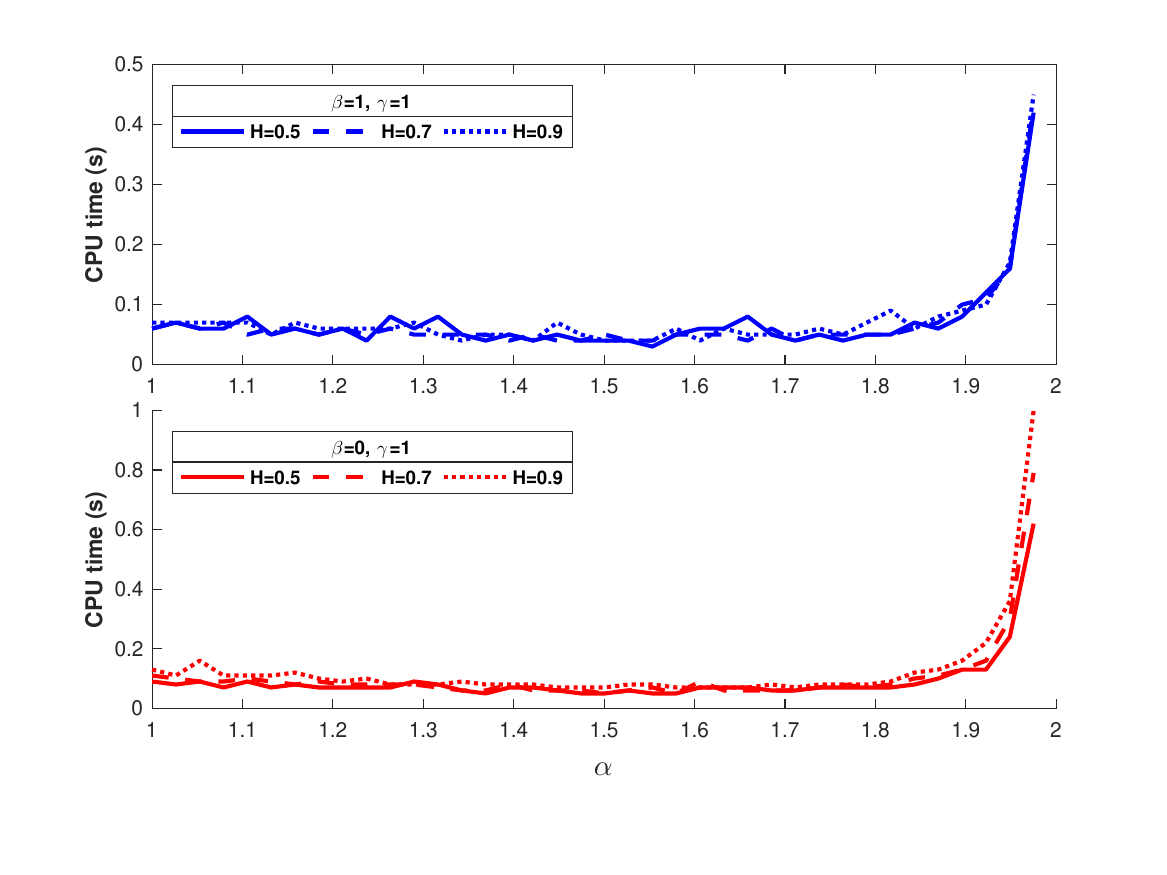}

}%
\subfloat[\label{fig:T=00003D1.5-1-1}T=1.5]%
{\includegraphics[width=0.5\textwidth]{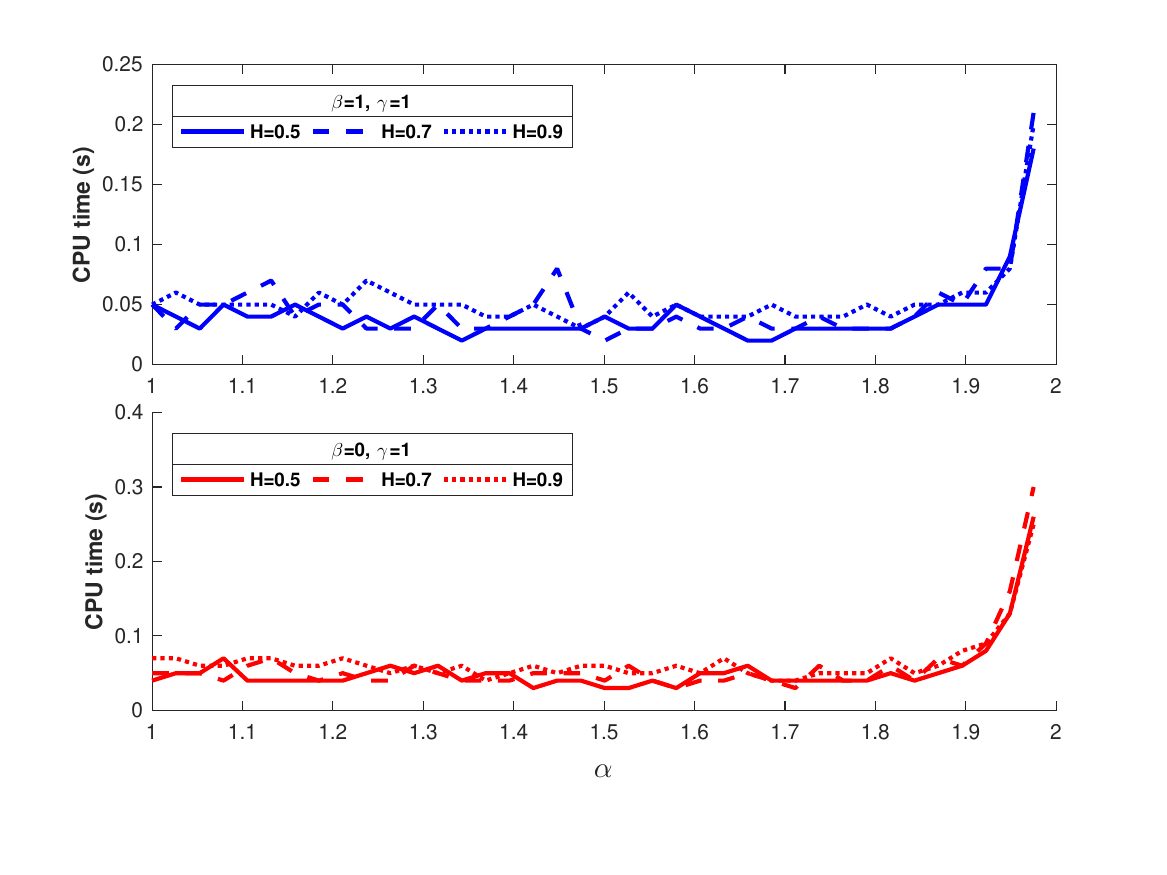}

}

\caption{\label{fig:Mixed-fractional-and-fractional-1}CPU times for the computation
of the Mixed-fractional and fractional CEV formula using $S_{0}=E=100$,
$r=5$\% and $\sigma=20$\%}
\end{figure}

Fig. (\ref{fig:Mixed-fractional-and-fractional-1}) addresses the
computational cost, measured as CPU time, of the mixed and pure fractional
models. For $\alpha\rightarrow2$, the formula becomes extremely expensive.
Nevertheless, there are no significative differences in terms of cost
between the classical and fractional approaches. 


\section{Greeks}

For to analyze the sensitivities of the pricing formula as function
of the parameters of the model, we compare the Greeks of both classical,
fractional and mixed-fractional CEV models.

The most common sensitivities are related to the price, maturity,
volatility and interest rate. We use the results given in Ref. \cite{larguinho2013computation}
for $\Delta,$ $\text{\ensuremath{\Gamma}}$, $\nu$, $\Theta$ and
$\rho$ Greeks under the classical CEV model and here we carefully
extent it to the fractional cases. 

\subsection{Delta}

\begin{eqnarray}
\Delta & = & \text{\ensuremath{\frac{\partial C}{\partial S}}}\nonumber \\
 & = & Q\left(2z,2+\frac{2}{2-\alpha},2y\right)+\frac{2y\left(2-\alpha\right)}{S}\left[Sf\left(2z;4+\frac{2}{2-\beta},2y\right)-E\text{e}^{-rT}f\left(2y;\frac{2}{2-\beta},2z\right)\right]\label{eq:delta}
\end{eqnarray}
\\

\noindent where $f$ is the non-central-$\chi^{2}$ PDF defined in
\ref{eq:f}. 

\begin{figure}
\subfloat[{\label{fig:deltaS}$S\in\left[80,120\right]$ and $\alpha=1.5$}]%
{\includegraphics[width=0.5\textwidth]{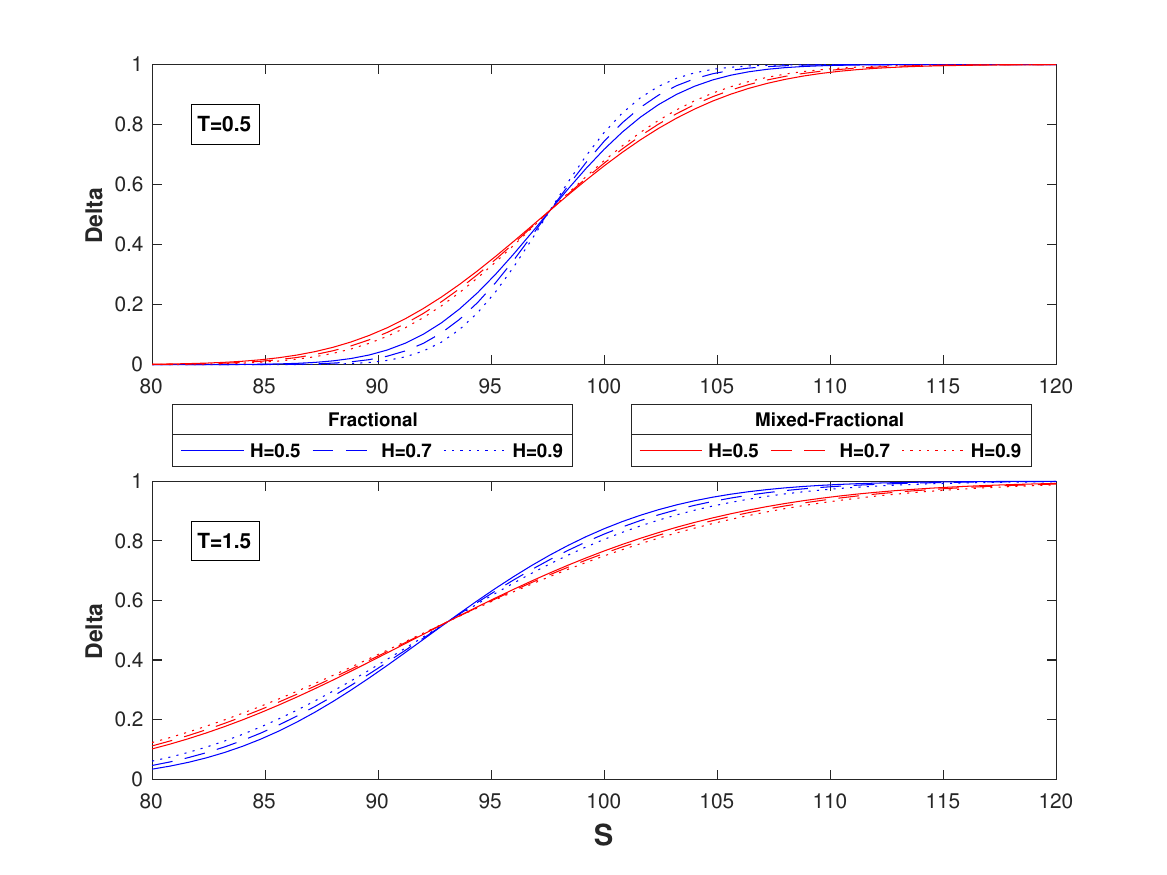}

}%
\subfloat[{\label{fig:deltab}S=100 and $\alpha\in\left[1,2\right]$}]%
{\includegraphics[width=0.5\textwidth]{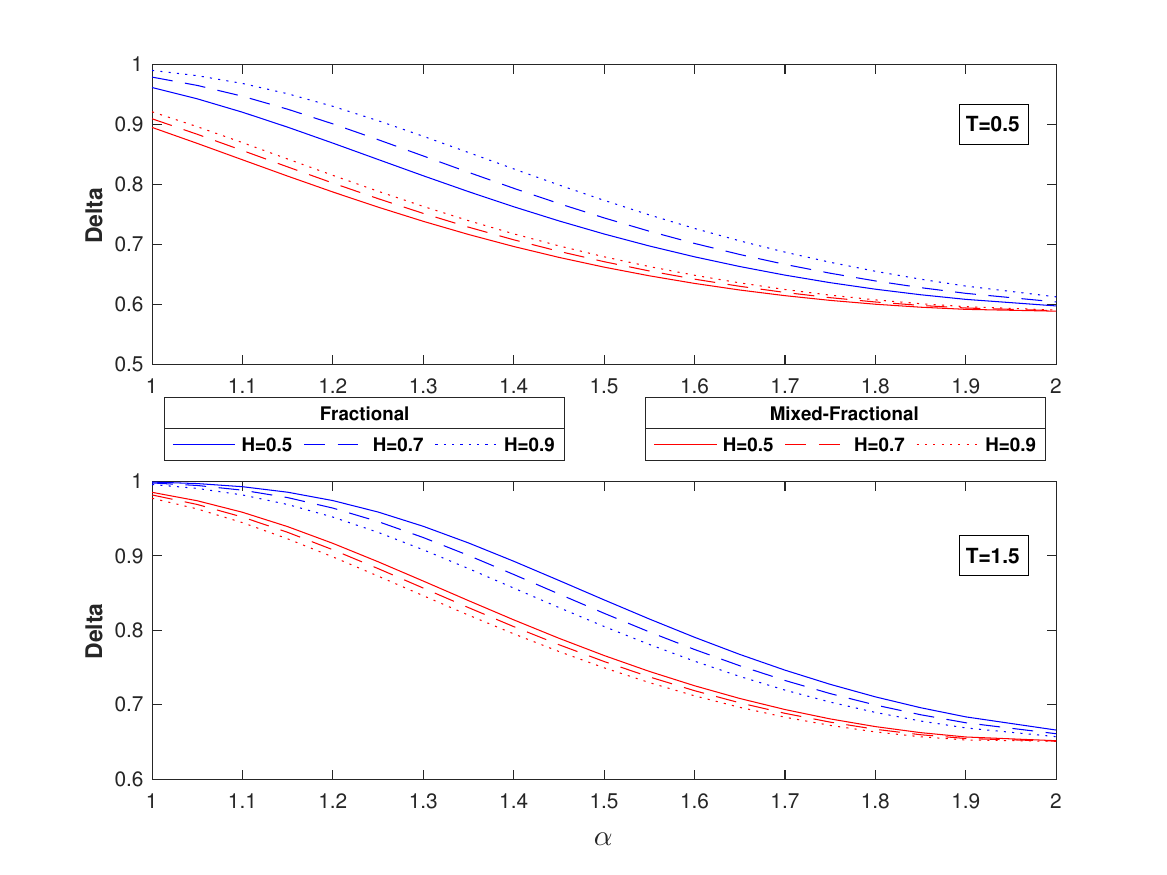}

}

\caption{\label{fig:Delta-for-fractional}Delta for fractional (blue) and mixed-fractional
(red) models setting $E$=100, $\sigma=20$\%, $r=5$\% and $\gamma=\beta=1$.}

\end{figure}

For the fractional case, we have:

\begin{eqnarray}
\Delta_{H} & = & \text{\ensuremath{\frac{\partial C_{H}}{\partial S}}}\nonumber \\
 & = & Q\left(2z_{H},2+\frac{2}{2-\alpha},2y_{H}\right)\nonumber \\
 &  & +\frac{2y_{H}\left(2-\alpha\right)}{S}\left[Sf\left(2z_{H};4+\frac{2}{2-\beta},2y_{H}\right)-E\text{e}^{-rT}f\left(2y_{H};\frac{2}{2-\beta},2z_{H}\right)\right]\label{eq:deltaH}
\end{eqnarray}
\\
\noindent and in the mixed-fractional:

\begin{eqnarray}
\Delta_{M} & = & \text{\ensuremath{\frac{\partial C_{M}}{\partial S}}}\nonumber \\
 & = & Q\left(2z_{M},2+\frac{2}{2-\alpha},2y_{M}\right)\nonumber \\
 &  & +\frac{2y_{M}\left(2-\alpha\right)}{S}\left[Sf\left(2z_{M};4+\frac{2}{2-\beta},2y_{M}\right)-E\text{e}^{-rT}f\left(2y_{M};\frac{2}{2-\beta},2z_{M}\right)\right]\label{eq:deltaM}
\end{eqnarray}
\\
The charts at the Figure \ref{fig:Delta-for-fractional} show the
behavior of the $\text{\ensuremath{\Delta},}\,\Delta_{H}$ and $\Delta_{M}$
varying the spot price (\ref{fig:deltaS}) and elasticity (\ref{fig:deltab})
for maturities below and above the unity. The solid blue line corresponds
to the classical CEV model.

\subsection{Gamma}

\begin{figure}
\subfloat[{$S\in\left[80,120\right]$ and $\alpha=1.5$}]%
{\includegraphics[width=0.5\textwidth]{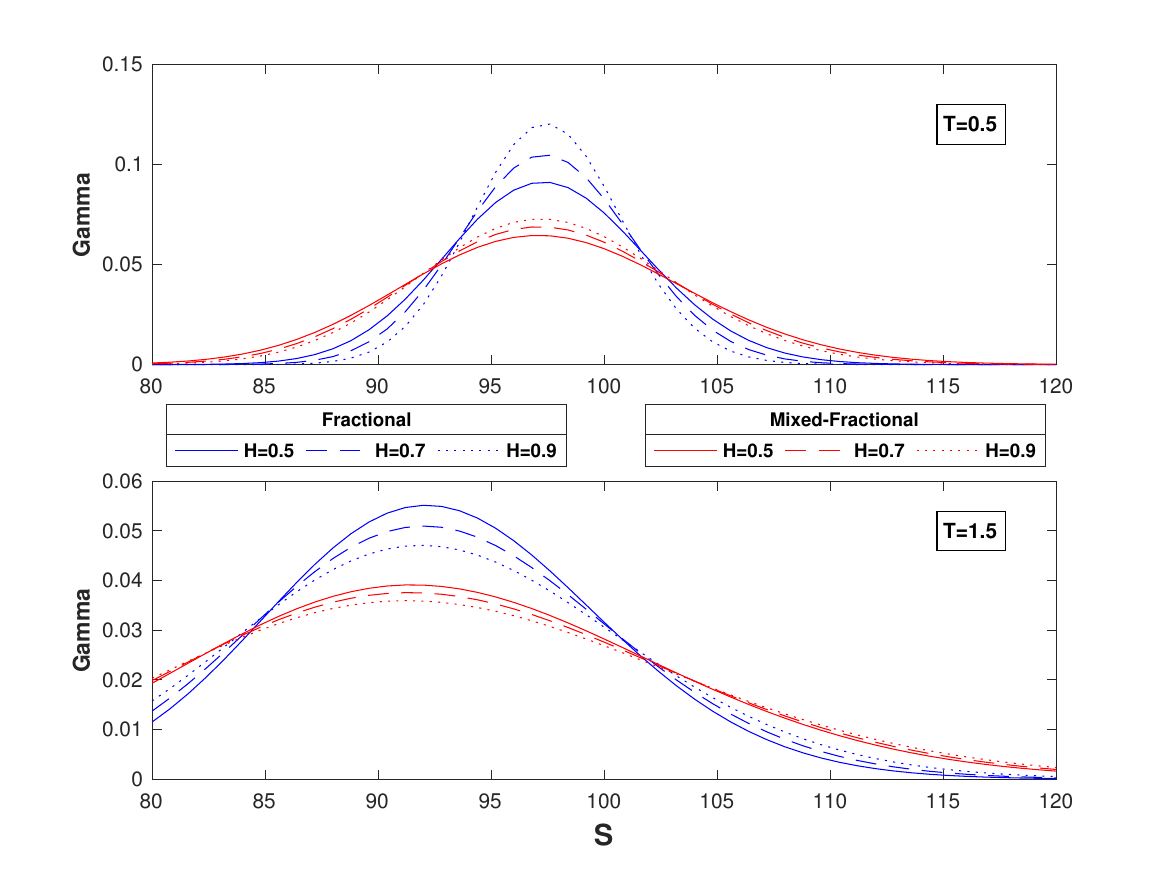}

}%
\subfloat[{S=100 and $\alpha\in\left[1,2\right]$}]%
{\includegraphics[width=0.5\textwidth]{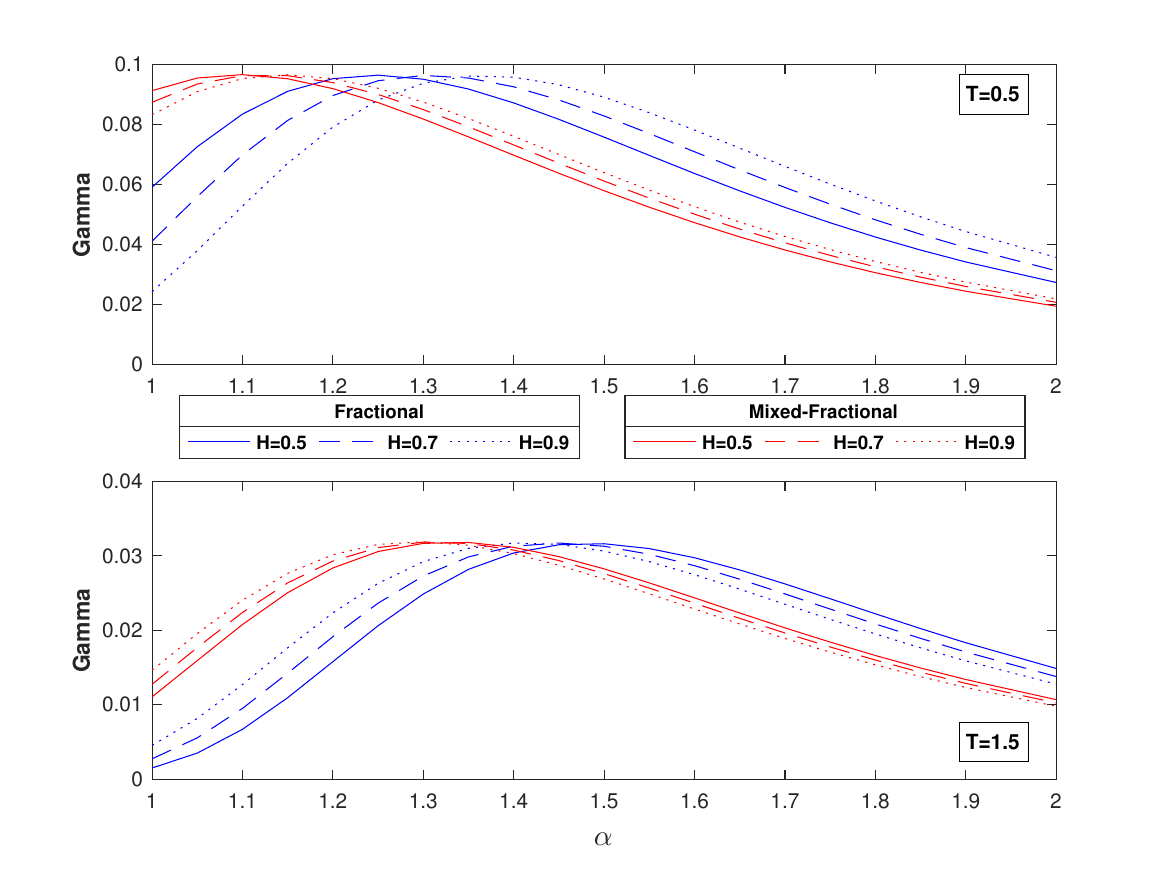}

}

\caption{\label{fig:Gamma-for-fractional}Gamma for fractional (blue) and mixed-fractional
(red) models setting $E$=100, $\sigma=20$\%, $r=5$\% and $\gamma=\beta=1$.}
\end{figure}

The eqs. \ref{eq:gamma}, \ref{eq:gammaH} and \ref{eq:gammaM} provide
the Gamma sensitivity for standard, fractional and mixed-fractional
CEV, respectively, plotting it at Fig. \ref{fig:Gamma-for-fractional}

\begin{eqnarray}
\Gamma & = & \text{\ensuremath{\frac{\partial^{2}C}{\partial S^{2}}}}\nonumber \\
 & = & \frac{\partial}{\partial S}\Delta\nonumber \\
 & = & \frac{2y\left(2-\alpha\right)^{2}}{S}\left\{ \left[\frac{\left(3-\alpha\right)}{2-\alpha}-y\right]f\left(2z;4+\frac{2}{2-\beta},2y\right)+yf\left(2z;6+\frac{2}{2-\beta},2y\right)\right\} \nonumber \\
 &  & +\frac{2y\left(2-\alpha\right)^{2}}{S^{2}}E\text{e}^{-rT}\left[yf\left(2y;\frac{2}{2-\beta},2z\right)-zf\left(2y;2+\frac{2}{2-\beta},2z\right)\right]\label{eq:gamma}
\end{eqnarray}
\\

\begin{eqnarray}
\Gamma_{H} & = & \text{\ensuremath{\frac{\partial^{2}C_{H}}{\partial S^{2}}}}\nonumber \\
 & = & \frac{2y_{H}\left(2-\alpha\right)^{2}}{S}\left\{ \left[\frac{\left(3-\alpha\right)}{2-\alpha}-y_{H}\right]f\left(2z_{H};4+\frac{2}{2-\beta},2y_{H}\right)+y_{H}f\left(2z_{H};6+\frac{2}{2-\beta},2y_{H}\right)\right\} \nonumber \\
 &  & +\frac{2y_{H}\left(2-\alpha\right)^{2}}{S^{2}}E\text{e}^{-rT}\left[y_{H}f\left(2y_{H};\frac{2}{2-\beta},2z_{H}\right)-z_{H}f\left(2y_{H};2+\frac{2}{2-\beta},2z_{H}\right)\right]\label{eq:gammaH}
\end{eqnarray}
\\

\begin{eqnarray}
\Gamma_{M} & = & \text{\ensuremath{\frac{\partial^{2}C_{M}}{\partial S^{2}}}}\nonumber \\
 & = & \frac{2y_{M}\left(2-\alpha\right)^{2}}{S}\left\{ \left[\frac{\left(3-\alpha\right)}{2-\alpha}-y_{M}\right]f\left(2z_{M};4+\frac{2}{2-\beta},2y_{M}\right)+y_{M}f\left(2z_{M};6+\frac{2}{2-\beta},2y_{M}\right)\right\} \nonumber \\
 &  & +\frac{2y_{M}\left(2-\alpha\right)^{2}}{S^{2}}E\text{e}^{-rT}\left[y_{M}f\left(2y_{M};\frac{2}{2-\beta},2z_{M}\right)-z_{M}f\left(2z_{M};2+\frac{2}{2-\beta},2y_{M}\right)\right]\label{eq:gammaM}
\end{eqnarray}

\subsection{Vega}

The partial derivative with respect to the volatility is called Vega
(commonly represented by the greek letter $\nu$). In the CEV model,
the volatility is a function of both $S$ and the parameter $\sigma$,
defined by $\boldsymbol{\sigma}^{2}=\sigma S^{\frac{\alpha-2}{2}}$.
Then the Vega for the CEV model is: 

\begin{eqnarray}
\nu & = & \frac{\partial C}{\partial\boldsymbol{\sigma}};\quad\sigma^{2}=\boldsymbol{\sigma}^{2}S^{2-\alpha}\nonumber \\
 & = & \frac{\partial C}{\partial\sigma}\frac{\partial\sigma}{\partial\boldsymbol{\sigma}}\nonumber \\
 & = & \left[\frac{\partial C}{\partial(2y)}\frac{\partial(2y)}{\partial k}\right]\frac{\left(-2k\right)}{\sigma}S^{\frac{2-\alpha}{2}}\nonumber \\
 & = & \frac{-4y}{\boldsymbol{\sigma}}\left[Sf\left(2z,4+\frac{2}{2-2\alpha},2y\right)-E\text{e}^{-rT}f\left(2y,\frac{2}{2-\alpha},2z\right)\right]\label{eq:Vega}
\end{eqnarray}

For the fractional and mixed-fractional models, we have:

\begin{eqnarray}
\nu_{H} & = & \frac{\partial C_{H}}{\partial\boldsymbol{\sigma}}\nonumber \\
 & = & \left[\frac{\partial C_{H}}{\partial(2y_{H})}\frac{\partial(2y_{H})}{\partial k_{H}}\right]\frac{\left(-2k_{H}\right)}{\sigma}S^{\frac{2-\alpha}{2}}\nonumber \\
 & = & \frac{-4y_{H}}{\boldsymbol{\sigma}}\left[Sf\left(2z_{H},4+\frac{2}{2-2\alpha},2y_{H}\right)-E\text{e}^{-rT}f\left(2y_{H},\frac{2}{2-\alpha},2z_{H}\right)\right]\label{eq:VegaH}
\end{eqnarray}

\begin{eqnarray}
\nu_{M} & = & \frac{\partial C_{M}}{\partial\boldsymbol{\sigma}}\nonumber \\
 & = & \left[\frac{\partial C_{M}}{\partial(2y_{M})}\frac{\partial(2y_{M})}{\partial k_{M}}\right]\frac{\left(-2k_{M}\right)}{\sigma}S^{\frac{2-\alpha}{2}}\nonumber \\
 & = & \frac{-4y_{M}}{\boldsymbol{\sigma}}\left[Sf\left(2z_{M},4+\frac{2}{2-2\alpha},2y_{M}\right)-E\text{e}^{-rT}f\left(2y_{M},\frac{2}{2-\alpha},2z_{M}\right)\right]\label{eq:VegaM}
\end{eqnarray}

\begin{figure}
\subfloat[{$\sigma\in\left[0.05,0.4\right]$ and $\alpha=1.5$}]%
{\includegraphics[width=0.5\textwidth]{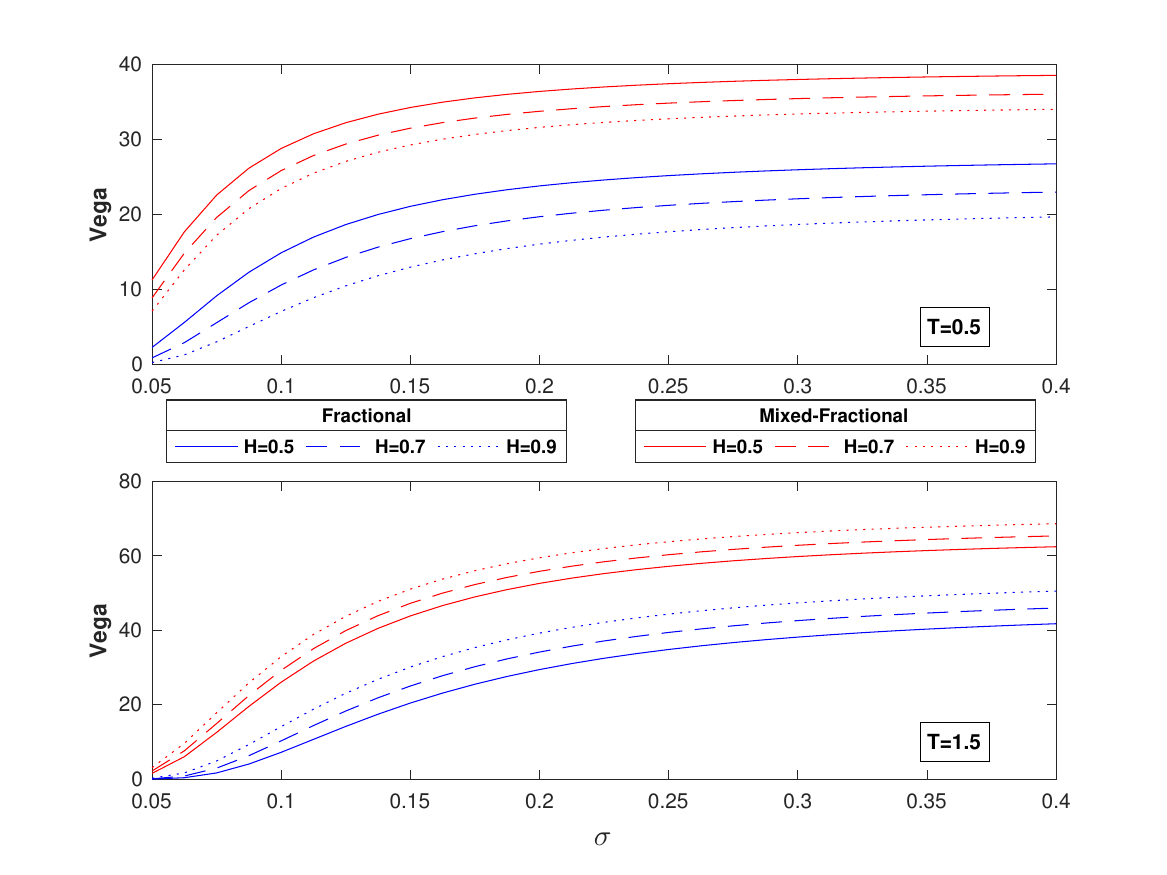}

}%
\subfloat[{$\sigma=20\%$ and $\alpha\in\left[1,2\right]$}]%
{\includegraphics[width=0.5\textwidth]{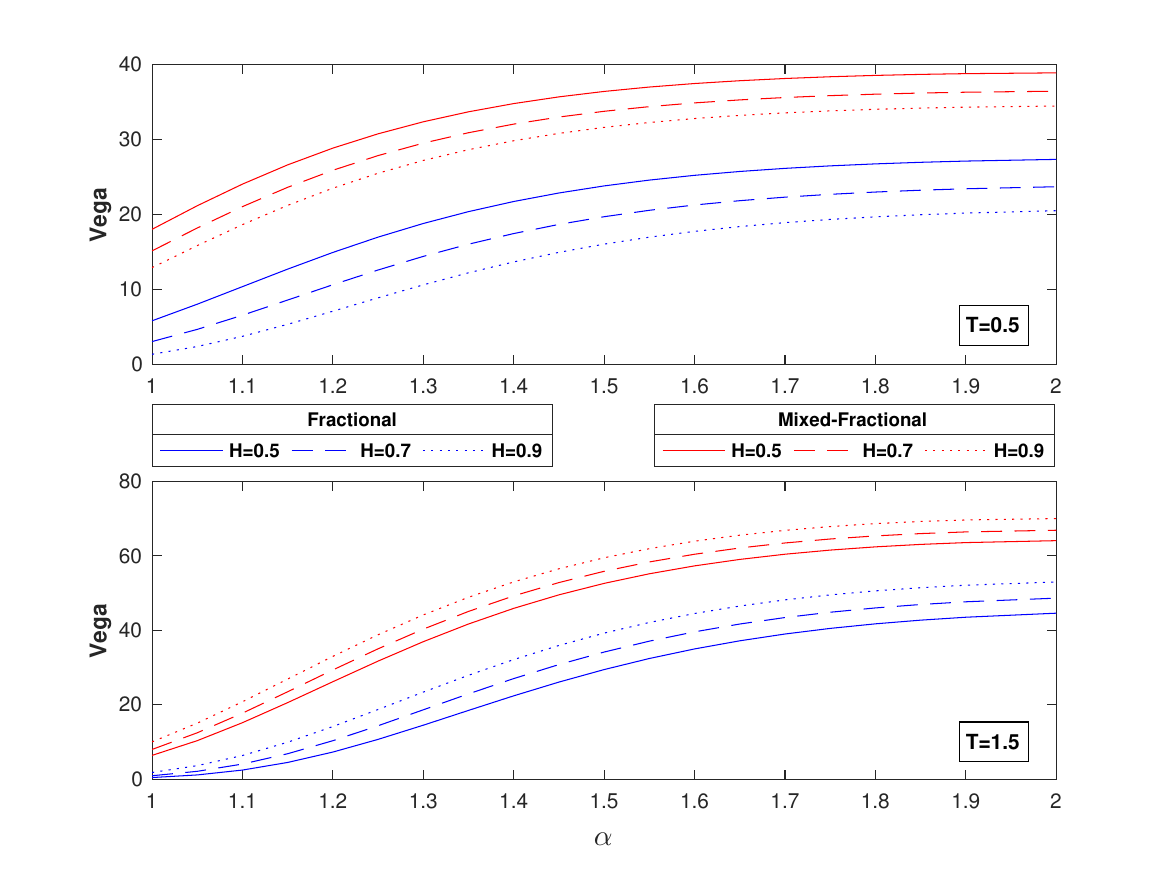}

}

\caption{\label{fig:Vega-for-fractional}Vega for fractional (blue) and mixed-fractional
(red) models setting $S=E$=100, $r=5$\% and $\gamma=\beta=1$.}
\end{figure}

The graphics in Fig. \ref{fig:Vega-for-fractional} display the Vega
for the fractional and mixed-fractional models. The standard CEV Vega
is plotted by the solid-blue line.

\subsection{Theta }

\begin{figure}
\subfloat[{$T\in\left[0.02,1.6\right]$ and $\alpha=1.5$\label{fig:thetaa}}]%
{\includegraphics[width=0.5\textwidth]{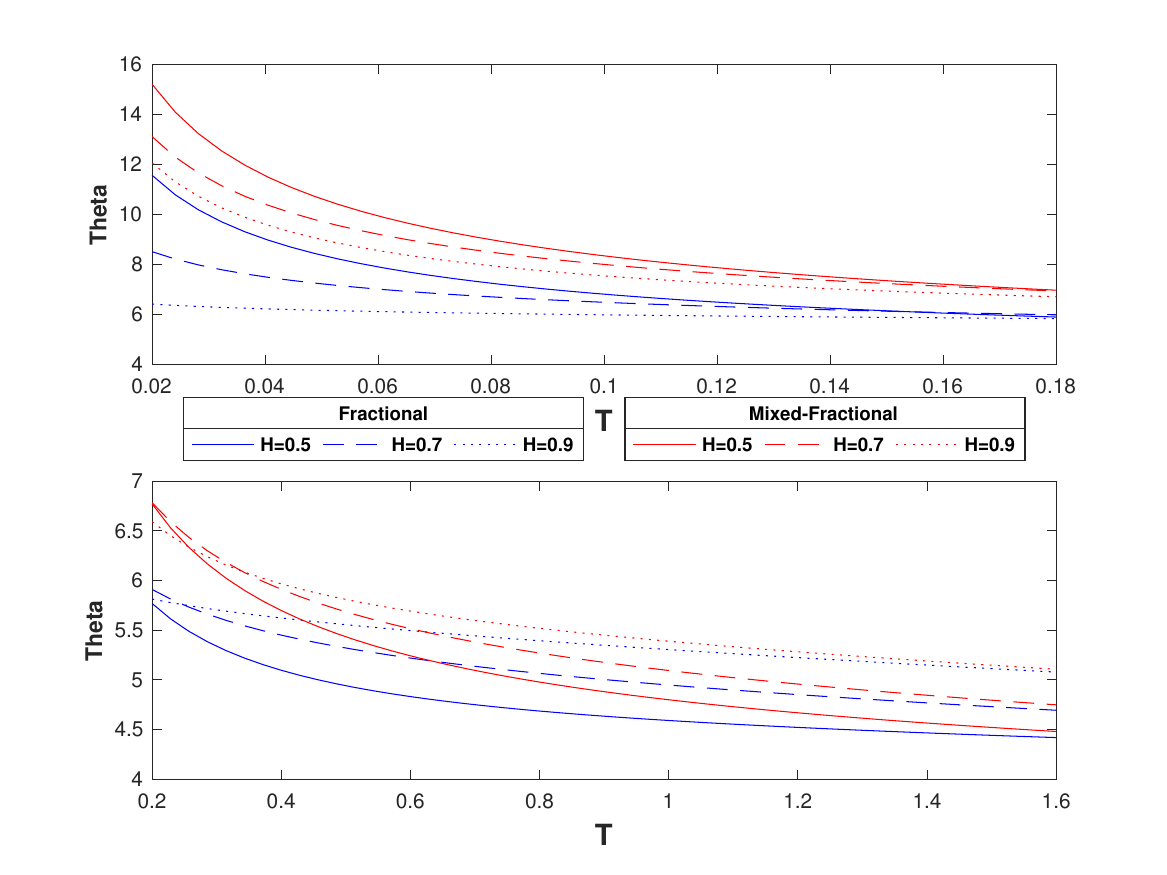}

}%
\subfloat[{\label{fig:thetab}$\alpha\in\left[1,2\right]$}]%
{\includegraphics[width=0.5\textwidth]{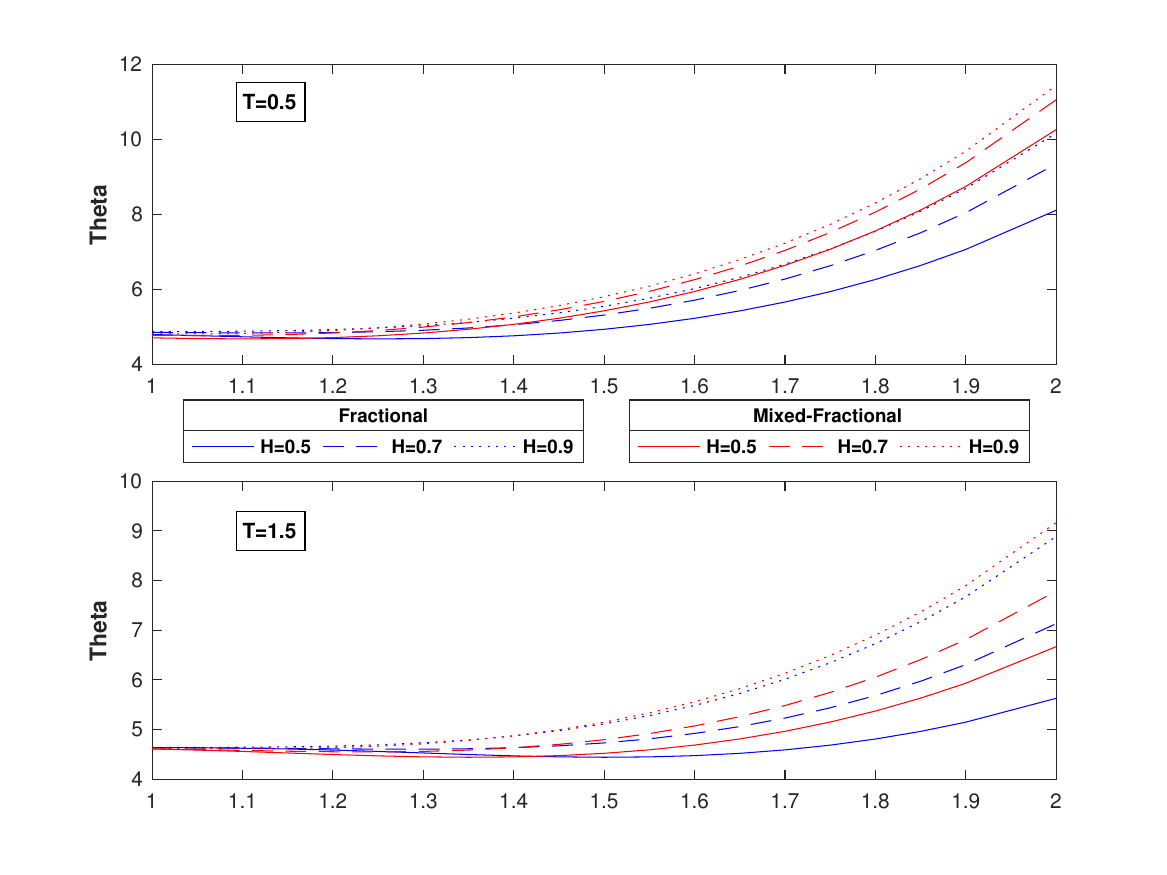}

}

\caption{\label{fig:Theta-for-fractional}Theta for fractional (blue) and mixed-fractional
(red) models setting $S=E$=100, $r=5$\%, $\sigma=$20\% and $\gamma=\beta=1$.}
\end{figure}

The change rate of the option price with respect to the maturity $T$
is noted by the greek $\Theta$, and is computed for the CEV model
at the Eq. \ref{eq:Theta}, for the fractional CEV at Eq. \ref{eq:ThetaH}
and for the mixed-fractional CEV at Eq. \ref{eq:ThetaM}. Fig. \ref{fig:Theta-for-fractional}
draw the shape of $\Theta_{H}$ and $\Theta_{M}$ under different
values of T (\ref{fig:thetaa}) and alpha (\ref{fig:thetab}).

\begin{eqnarray}
\Theta & = & \frac{\partial C}{\partial T}\nonumber \\
 & = & S\frac{\partial Q\left(2z,2+2/(2-\alpha),2y\right)}{\partial(2y)}\frac{\partial(2y)}{\partial T}+rE\text{e}^{-rT}\left[1-Q\left(2y,2/(2-\alpha),2z\right)\right]\nonumber \\
 &  & +E\text{e}^{-rT}\frac{\partial Q\left(2y,2/(2-\alpha),2z\right)}{\partial(2y)}\frac{\partial(2y)}{\partial T}\nonumber \\
 & = & -\frac{2yr\left(2-\alpha\right)}{\text{e}^{r\left(2-\alpha\right)T}-1}\left[Sf\left(2z,4+\frac{2}{2-\alpha},2y\right)-E\text{e}^{-rT}f\left(2y,\frac{2}{2-\alpha},2z\right)\right]\nonumber \\
 &  & +rE\text{e}^{-rT}\left[1-Q\left(2y,2/(2-\alpha),2z\right)\right]\label{eq:Theta}
\end{eqnarray}
\\

\begin{eqnarray}
\Theta_{H} & = & \frac{\partial C_{H}}{\partial T}\nonumber \\
 & = & S\frac{\partial Q\left(2z_{H},2+2/(2-\alpha),2y_{H}\right)}{\partial(2y_{H})}\frac{\partial(2y_{H})}{\partial T}+rE\text{e}^{-rT}\left[1-Q\left(2y_{H},2/(2-\alpha),2z_{H}\right)\right]\nonumber \\
 &  & +E\text{e}^{-rT}\frac{\partial Q\left(2y_{H},2/(2-\alpha),2z_{H}\right)}{\partial(2y_{H})}\frac{\partial(2y_{H})}{\partial T}\nonumber \\
 & = & -2y_{H}HT^{2H-1}\frac{\sigma^{2}\left(2-\alpha\right)^{2}}{\phi(T)}\left[Sf\left(2z_{H},4+\frac{2}{2-\alpha},2y_{H}\right)-E\text{e}^{-rT}f\left(2y_{H},\frac{2}{2-\alpha},2z_{H}\right)\right]\nonumber \\
 &  & +rE\text{e}^{-rT}\left[1-Q\left(2y_{H},2/(2-\alpha),2z_{H}\right)\right]\label{eq:ThetaH}
\end{eqnarray}
\\
\begin{eqnarray}
\Theta_{M} & = & \frac{\partial C_{M}}{\partial T}\nonumber \\
 & = & -2y_{M}\frac{\sigma^{2}\left(2-\alpha\right)^{2}}{\phi'(T)}\left[\gamma HT^{2H-1}+\frac{\beta}{2}\right]\left[Sf\left(2z_{M},4+\frac{2}{2-\alpha},2y_{M}\right)\right.\nonumber \\
 &  & \left.-E\text{e}^{-rT}f\left(2y_{M},\frac{2}{2-\alpha},2z_{M}\right)\right]+rE\text{e}^{-rT}\left[1-Q\left(2y_{M},2/(2-\alpha),2z_{M}\right)\right]\label{eq:ThetaM}
\end{eqnarray}
\\

\subsection{Rho}

\begin{figure}
\subfloat[{$r\in\left[0.01,0.2\right]$ and $\alpha=1.5$}]%
{\includegraphics[width=0.5\textwidth]{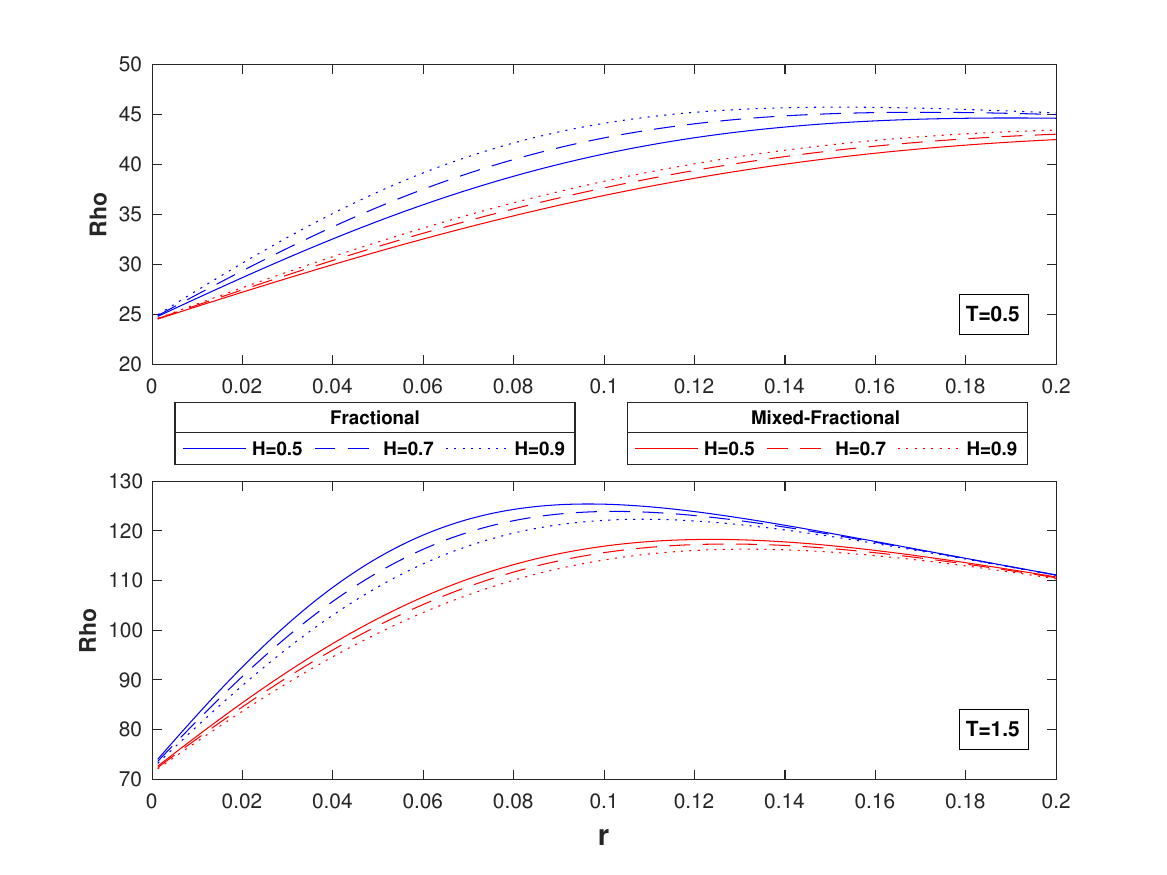}

}%
\subfloat[{$r=5$\% and $\alpha\in\left[1,2\right]$}]%
{\includegraphics[width=0.5\textwidth]{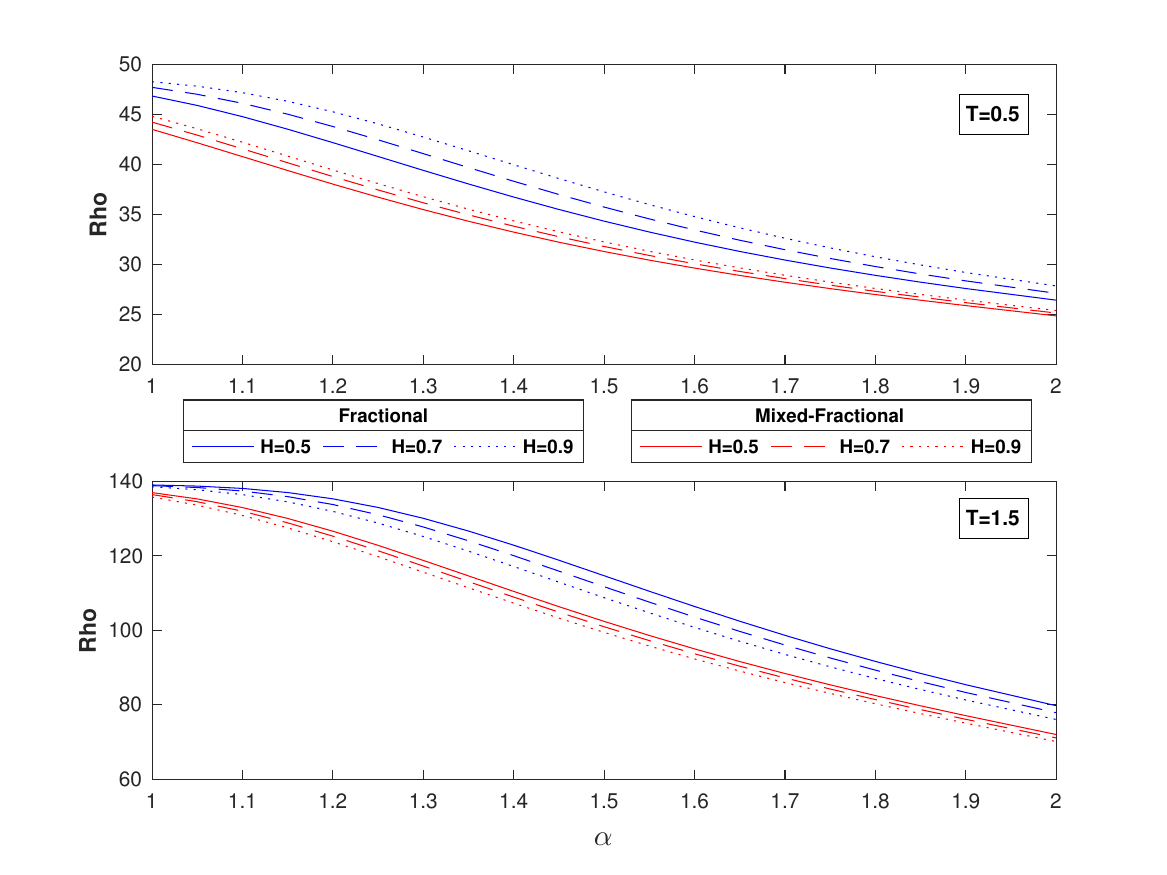}

}

\caption{\label{fig:Rho}Rho for fractional (blue) and mixed-fractional (red)
models setting $S=E$=100, $\sigma=20$\% and $\gamma=\beta=1$.}
\end{figure}

Finally, the sensitivity respect to the risk-free-interest rate, for
the CEV model and its extensions (fractional and mixed-fractional),
are shown at Fig. \ref{fig:Rho} and explicitly computed at the eqs.
\ref{eq:rho}-\ref{eq:rhoH}-\ref{eq:rhoM}.

\begin{eqnarray}
\rho & = & \frac{\partial C}{\partial r}\nonumber \\
 & = & S\frac{\partial Q\left(2z,2+2/(2-\alpha),2y\right)}{\partial(2y)}\frac{\partial(2y)}{\partial r}+TE\text{e}^{-rT}\left[1-Q\left(2y,2/(2-\alpha),2z\right)\right]\nonumber \\
 &  & +E\text{e}^{-rT}\frac{\partial Q\left(2y,2/(2-\alpha),2z\right)}{\partial(2y)}\frac{\partial(2y)}{\partial T}\nonumber \\
 & = & 2y\left[\frac{1}{r}-\frac{\left(2-\alpha\right)T}{\text{e}^{r\left(2-\alpha\right)T}-1}\right]\left[Sf\left(2z,4+\frac{2}{2-\alpha},2y\right)-E\text{e}^{-rT}\left(2y,\frac{2}{2-\alpha},2z\right)\right]\nonumber \\
 &  & +TE\text{e}^{-rT}\left[1-Q\left(2y,2/(2-\alpha),2z\right)\right]\label{eq:rho}
\end{eqnarray}
\\
\begin{eqnarray}
\rho_{H} & = & \frac{\partial C_{H}}{\partial r}\nonumber \\
 & = & S\frac{\partial Q\left(2z_{H},2+2/(2-\alpha),2y_{H}\right)}{\partial(2y_{H})}\frac{\partial(2y)}{\partial r}+TE\text{e}^{-rT}\left[1-Q\left(2y_{H},2/(2-\alpha),2z_{H}\right)\right]\nonumber \\
 &  & +E\text{e}^{-rT}\frac{\partial Q\left(2y_{H},2/(2-\alpha),2z_{H}\right)}{\partial(2y_{H})}\frac{\partial(2y_{H})}{\partial T}\nonumber \\
 & = & 2y_{H}\left[\frac{2H}{r}-\frac{H\left(2-\alpha\right)^{2}\sigma^{2}T^{2H}}{r\phi(T)}\right]\left[Sf\left(2z_{H},4+\frac{2}{2-\alpha},2y_{H}\right)-E\text{e}^{-rT}\left(2y_{H},\frac{2}{2-\alpha},2z_{H}\right)\right]\nonumber \\
 &  & +TE\text{e}^{-rT}\left[1-Q\left(2y_{H},2/(2-\alpha),2z_{H}\right)\right]\label{eq:rhoH}
\end{eqnarray}
\\
\begin{eqnarray}
\rho_{M} & = & \frac{\partial C_{M}}{\partial r}\nonumber \\
 & = & \frac{2y_{M}}{\phi'(T)r}\left\{ \gamma\left[2H\phi(T)-H\left(2-\alpha\right)^{2}\sigma^{2}T^{2H}\right]+\beta\frac{\left(2-\alpha\right)\sigma^{2}}{2r}\left[\text{e}^{r\left(2-\alpha\right)T}-1-r\left(2-\alpha\right)T\right]\right\} \nonumber \\
 &  & \times\left[Sf\left(2z_{M},4+\frac{2}{2-\alpha},2y_{M}\right)-E\text{e}^{-rT}\left(2y_{M},\frac{2}{2-\alpha},2z_{M}\right)\right]\nonumber \\
 &  & +TE\text{e}^{-rT}\left[1-Q\left(2y_{M},2/(2-\alpha),2z_{M}\right)\right]\label{eq:rhoM}
\end{eqnarray}

\section{Summary}

In this paper, the constant elasticity of variance model is studied,
adding a fractal feature to the Brownian motion to address the long-memory
in financial markets. Besides, to deal with the non-arbitrage issue
under pure fractional regimes, a mixed fractional CEV model is developed. 

Then, using the fractional Itô calculus and the fractional generalization
of the Fokker-Planck equations, an analytical and compact option pricing
scheme for a European Call, based on the complementary non-central-chi-squared
density function and the M-Whittaker function, is provided for both
approaches. The convergence to both the classical CEV model and the
fractional \& mixed-fractional Black Scholes formula are shown for
the limit cases $H\rightarrow1/2$ and $\alpha\rightarrow2$, respectively,
and then the proposed fractional extensions could be interpreted as
a generalization of the classical CEV and Black-Scholes model. Besides,
the Greeks are computed showing their behavior under different values
of the Hurst exponent, considering maturities lower and greater than
one.

Since the added terms on the call formula in both fractional models,
in relation to the classical CEV, doesn't have a dependency on the
strike price, the fractional and mixed-fractional CEV keep the capability
 to address the smile-skew issue.

\section{Acknowledgements}

I thank FIAS for financial support and both Dr. Nils Bertschinger
and Dr. Marcelo Villena for helpful discussions.

\bibliographystyle{unsrt}
\bibliography{19_Users_axel_Desktop_fBM_fBM}

\appendix

\section{\label{sec:Risk-neutral-pricing-in}Risk-neutral pricing in the fractional
CEV model}

As pointed in \cite{hu2003fractional,necula2002option}, the fractional
Clark-Ocone theorem and quasi-expectations are used for to price a
derivative under fractional Brownian motion. Here, we use the derivation
of Hu and \O skendal \cite{hu2003fractional} for to obtain a risk-neutral
pricing at time $t=0$, but using a fractional CEV approach instead
a fractional GBM.

Since the market is complete, at time $t$, a derivative $F$ is replicated
by:

\begin{equation}
F(t)=a(t)B(t)+b(t)S(t)\label{eq:Replication}
\end{equation}
\\
\noindent where $a$ and $b$ are weights, $B$ is a money bank account
(bond) which pays a continuously composed interest rate $r$ (risk-less
interest rate; i.e, $\text{d}B(t)=rB(t)\text{d}t$) and $S$ is ruled
by the Eq. \ref{eq:fCEV}.

Later:

\begin{eqnarray}
\text{d}F(t) & = & a(t)\text{d}B(t)+b(t)\text{d}S(t)\nonumber \\
 & = & a(t)rB(t)\text{d}t+b(t)\left[rS(t)\text{d}t+\sigma\left(S(t)\right)^{\alpha/2}\text{d}B_{t}^{H}\right]\nonumber \\
 & = & r\left[a(t)B(t)+b(t)S(t)\right]\text{d}t+b(t)\sigma\left(S(t)\right)^{\alpha/2}\text{d}B_{t}^{H}\nonumber \\
 & = & rF(t)\text{d}t+b(t)\sigma S(t)\text{d}B_{t}^{H}\label{eq:dF}
\end{eqnarray}
\\

Multiplying \ref{eq:dF} by $\text{e}^{-rt}$ and integrating it from
zero to $t$, we get:

\begin{equation}
\text{e}^{-rt}F(t)=F(0)+\int_{0}^{t}\text{e}^{-rt}b(t)\sigma\left(S(t)\right)^{\alpha/2}\text{d}B_{t}^{H}\label{eq:eDf}
\end{equation}
\\

On the other hand, the Clark-Ocone theorem for standard Brownian motions
is given by \cite{ocone1984malliavin}:

\[
G(t)=\mathbb{E}\left[G(T)\right]+\int_{0}^{T}D_{t}\mathbb{E}\left[G(t)|\mathcal{F}_{t}\right]\text{d}B_{t}
\]
\\

\noindent where $D_{t}$ is the Malliavin derivative and $\mathcal{F}_{t}$
is the natural filtration of Brownian motion. The fractional extension
of the Eq. is provided by Refs. \cite{hu2003fractional,bender2003clark}:

\begin{equation}
G(t)=\mathbb{E}\left[G(T)\right]+\int_{0}^{T}D_{t}\mathbb{E}\left[G(t)|\mathcal{F}_{t}^{H}\right]\text{d}B_{t}^{H}\label{eq:C-O-F}
\end{equation}
\\

\noindent being $\mathcal{F}_{t}^{H}$ the $\sigma$-algebra generated
by $B_{s}^{H}$, $s\leq t$. Put $G(t)=\text{e}^{-rt}F(t)$ in \ref{eq:C-O-F}:

\begin{equation}
\text{e}^{-rt}F(t)=\mathbb{E}\left[\text{e}^{-rt}F(T)\right]+\int_{0}^{T}D_{t}\mathbb{E}\left[\text{e}^{-rt}F(t)|\mathcal{F}_{t}^{H}\right]\text{d}B_{t}^{H}\label{eq:fractionalCO}
\end{equation}
\\

Comparing the expressions \ref{eq:eDf}-\ref{eq:fractionalCO} we
arrive to the completeness of the market by:

\[
D_{t}\mathbb{E}\left[F(t)|\mathcal{F}_{t}^{H}\right]=b(t)\sigma\left(S(t)\right)^{\alpha/2}
\]
\\

\noindent and

\begin{equation}
F(0)=\mathbb{E}\left[\text{e}^{-rt}F(T)\right]\label{eq:discounted}
\end{equation}
\\

Then, at the initial time, the price of a derivative is computed discounted
the expected value, as in the classical model driven by a Brownian
motion.

For the mixed-fractional case, the extension is straightforward and
is shown in \cite{sun2013pricing,feng2010pricing} for the mixed GBM. 
\end{document}